\documentclass[aps,pra,twocolumn,showpacs,superscriptaddress]{revtex4}
  
\usepackage{graphicx}
\usepackage{epstopdf}
\usepackage{amsmath}
\usepackage{amssymb}
\usepackage{verbatim}

\input{epsf}
\begin{document}

\title{Thermodynamics of the two-component Fermi gas with unequal masses at unitarity}

\author{K.~M. Daily}
\affiliation{Department of Physics and Astronomy,
Washington State University,
  Pullman, Washington 99164-2814, USA}
\author{D. Blume}
\affiliation{Department of Physics and Astronomy,
Washington State University,
  Pullman, Washington 99164-2814, USA}

\date{\today}

\begin{abstract}
We consider mass-imbalanced two-component Fermi gases for which 
the unequal-mass atoms interact
via a zero-range model potential with a diverging $s$-wave scattering 
length $a_s$, i.e., with $1/a_s=0$. 
The high temperature thermodynamics of the
harmonically trapped and homogeneous systems 
are examined using a virial expansion approach up to third order in the 
fugacity.
We find that the universal part of the third-order 
virial coefficient associated with two light atoms and one heavy atom is
negative, while that associated with two heavy and one light atom 
changes sign from negative to positive as
the mass ratio $\kappa$ increases, and diverges 
when Efimov physics sets in at $\kappa=13.61$.
By examining the Helmholtz free energy, we find that the
equilibrium polarization of the trapped and homogeneous systems is 0 for 
$\kappa=1$, but finite for $\kappa \ne 1$ (with a majority of heavy particles).
Compared to the equilibrium polarization of the non-interacting system, 
the equilibrium polarization at unitarity is increased for the trapped
system and decreased for the homogeneous system.
We find that unequal-mass Fermi gases are stable for all polarizations.
\end{abstract}


\maketitle

\section{Introduction}
\label{sec_intro}
It has been predicted that ultracold two-component Fermi gases with
mismatching Fermi surfaces support novel phases such as the 
Fulde-Ferrell-Larkin-Ovchinnikov~\cite{FuldeFerrell1964,LarkinOvchinnikov1964,LarkinOvchinnikov1965} state 
or other interesting phase separations, both in the 
homogeneous and inhomogeneous systems~\cite{RevModPhys2008,CLoboPRL2006,DESheehyPRL2006,ABulgacPRA2007,ARecatiPRA2008,PilatiGiorginiPRL2008}.
In equal-mass systems, the two Fermi surfaces differ if the number of
fermions of species 1 and 2 differ, i.e., if the system exhibits a
spin imbalance~\cite{YShinPRL2006,GBPartridgePRL2006}.  
Alternatively, mismatching Fermi surfaces arise if
species 1 and 2 have different masses~\cite{MIskinPRL2006,STWuPRB2006,CHPaoPRA2007,MMParishPRL2007,MIskinArxiv2008,CarlsonPRL2009,IBausmerthPRA2009,JStecherPRA2007}.
Experimentally, mass-imbalanced systems can be realized by simultaneously
trapping, e.g., $\!^6$Li and $\!^{40}$K~\cite{EWillePRL2008,MTaglieberPRL2008,FMSpiegelhalderPRA2010,LCostaPRL2010,ARidingerEPJD2011,DNaikEPJD2011,ATrenkwalderPRL2011}.
The application of an external magnetic field in the vicinity of a 
Fano-Feshbach resonance allows for the realization of strongly-interacting
gases, thus motivating our studies of the unequal-mass Fermi gas at unitarity.

At the few-body level, two-component 
unequal mass systems with short-range interspecies interactions and
mass ratio $\kappa$
are interesting because they support a variety of intriguing states~\cite{DBlumeReview2011}.
At unitarity, the three-body system consisting of two heavy and one light
particle  [the $(n_1,n_2)=(2,1)$ system] 
with zero-range interspecies interactions supports no three-body
bound state in free space 
for $\kappa \le 8.62$.  For $8.62 \le \kappa \le 13.61$, however,
the $(n_1,n_2)=(2,1)$ 
system can support a bound state whose binding energy is determined
by a three-body parameter~\cite{Petrov2003,WernerCastinPRA2006,NishidaSonTan2008,DBlumePRL2010,DBlumePRA2010}.
For yet larger mass ratios ($\kappa > 13.61$),
the three-body system in free space supports an infinite number of three-body
Efimov states, which are geometrically spaced~\cite{VNEfimov1970,EfimovLetter1972,Efimov1973}.
The spacing depends on the mass ratio $\kappa$ between the heavy and light
particles with the energy of the most deeply bound state being determined
by the so-called three-body Efimov parameter~\cite{EBraatenPhysRep2006}.

The objective of this paper is to investigate the thermodynamics
of inhomogeneous and homogeneous two-component Fermi gases
as a function of the mass ratio $\kappa$.
Throughout, we consider the unitary regime where the interspecies
$s$-wave scattering length diverges and determine the system properties
as functions of the polarization $P$, $P=(N_1-N_2)/(N_1+N_2)$, and the
temperature $T$. We are limiting ourselves to the so-called high temperature
regime, where the virial equation of state up to the third order is expected
to provide a valid description~\cite{HoPRL2004,HoMuellerPRL2004}.
For the trapped equal mass system, 
the virial equation of state up to the third order
has been shown to be applicable over the temperature range of well above
the Fermi temperature $T_F$ down to about $T_F/2$~\cite{LiuPRL2009,HuNewJPhys2010,LiuPRA2010,TokyoScience2010,SalomonNature2010}.

Our key findings are: (i) The second order virial equation of state
in a harmonic trap is independent of the mass ratio $\kappa$.  The
mass ratio first enters at the third order in the virial expansion.
(ii) The equilibrium state of the trapped system at unitarity
is stable and favors a majority of heavy particles for $\kappa \ne 1$.
(iii) The equation of state of the homogeneous system
depends strongly on the mass ratio. The mass
ratio first enters at the zeroth order in the virial expansion.
(iv) The equilibrium state of the homogeneous system at unitarity
is stable and favors a
majority of heavy particles.
The equilibrium polarization exhibits an intricate dependence on the
third-order virial coefficients.

The paper is organized as follows.  Section~\ref{sec_virials} defines the
virial expansion and the first few virial expansion coefficients
for a two-component Fermi gas in a harmonic trap where both species feel the
same angular trapping frequency $\omega$.
The third order virial expansion coefficients at unitarity are determined as
functions of the temperature and the mass ratio.
Sections~\ref{sec_thermo2} and~\ref{sec_thermo3}
explore the virial equation of state of the
harmonically trapped and homogeneous two-component Fermi gas at unitarity.
Finally, Sec.~\ref{sec_conclusion} concludes. 
Appendix~\ref{sec_thermo1} summarizes the virial expansion for 
the harmonically trapped single-component Fermi gas
and connects the results with those for the two-component Fermi gas.

\section{Determination of the virial coefficients}
\label{sec_virials} 
The model Hamiltonian $H(n_1,n_2)$ 
that describes unequal mass two-component Fermi gases
with $n_1$ heavy and $n_2$ light particles in a harmonic trap
with angular trapping frequency $\omega$ reads
\begin{align}
\label{eq_Hamiltonian}
& H(n_1,n_2) = \sum_{j=1}^{n_1} \left( \frac{-\hbar^2}{2 m_1} 
\vec{\nabla}_{\vec{r}_j}^2 + 
\frac{1}{2} m_1 \omega^2 \vec{r}_j^2 \right) \\
& + \sum_{j=n_1+1}^{n_1+n_2} \left( \frac{-\hbar^2}{2 m_2} 
\vec{\nabla}_{\vec{r}_j}^2 +
\frac{1}{2} m_2 \omega^2 \vec{r}_j^2 \right) \nonumber
+ \sum_{i=1}^{n_1} \sum_{j=n_1+1}^{n_1+n_2} V_{tb}(r_{ij}),
\end{align}
where $m_1$ denotes the mass of the heavy species and $m_2$ denotes
the mass of the light species.
We define the mass ratio $\kappa$ as $\kappa=m_1/m_2$ ($\kappa \ge 1$).
In Eq.~(\ref{eq_Hamiltonian}), $\vec{r}_j$ denotes the
position vector of the $j^{th}\!$ atom measured with respect 
to the trap center.
The interspecies interaction potential $V_{tb}$ is parameterized by a
zero-range potential,
$V_{tb}(r_{ij})=(2\pi\hbar^2 a_s/ \mu_{red}) 
\delta(\vec{r}_{ij})(\partial/\partial r_{ij}) r_{ij}$,
where $\mu_{red}$ denotes the reduced mass, $\mu_{red}=m_1m_2/(m_1+m_2)$,
$a_s$ the interspecies $s$-wave scattering length, and
$r_{ij}$ the interparticle distance, $r_{ij}=|\vec{r}_i-\vec{r}_j|$.

If the complete energy spectrum of the Hamiltonian given in 
Eq.~(\ref{eq_Hamiltonian}) were known,
one could calculate thermodynamic quantities like the
average energy $U$ or the entropy $S$ from the
free energy $F$, $F=-k_B T \ln{Q_{n_1,n_2}}$, 
using the canonical ensemble.  Here,
$k_B$ is Boltzmann's constant and $Q_{n_1,n_2}$ is
the canonical partition function~\cite{McQuarrie,HuangStatMech},
\begin{align}
\label{eq_canonicalPartFcta}
Q_{n_1,n_2} = \; \mbox{Tr} \; \exp{[-H(n_1,n_2)/(k_B T)]},
\end{align}
where Tr is the trace operator. To evaluate the trace, we insert
a complete set of eigenstates, yielding
\begin{align}
\label{eq_canonicalPartFctb}
Q_{n_1,n_2} = \sum_j \exp[-E_j^{(n_1,n_2)}/(k_B T)].
\end{align}
The summation index $j$ 
collectively denotes the complete set of quantum numbers allowed by symmetry.

In practice, it is easier to work in the grand canonical
ensemble, which can be considered to be a collection of canonical ensembles
with $n_1$ and $n_2$ particles of species 1 and 2
in thermal equilibrium with each other.
In the grand canonical ensemble, 
the chemical potentials $\mu_1$ and $\mu_2$ of the two species are fixed.
This implies that the system is characterized by the average numbers
$N_1$ and $N_2$ of atoms of species 1 and 2.
The thermodynamic potential $\Omega^{(2)}$ of the two-component Fermi
gas in the grand canonical ensemble is~\cite{HuangStatMech}
\begin{align}
\label{eq_thermPot1a}
\Omega^{(2)} = & -k_B T \ln \mbox{Tr} \; \exp[-(H(n_1,n_2) \nonumber \\ 
&- \mu_1 n_1 - \mu_2 n_2)/(k_B T)],
\end{align}
where the trace operator now extends over $n_1$ and $n_2$.
Rewritten in terms of the fugacities $z_i$, $z_i=\exp[\mu_i/(k_B T)]$,
the thermodynamic potential reads
\begin{align}
\label{eq_thermPot1b}
\Omega^{(2)} & = -k_B T \ln \left[ \sum_{n_1=0}^{\infty} 
  \sum_{n_2=0}^{\infty} Q_{n_1,n_2} z_1^{n_1} z_2^{n_2} \right].
\end{align}

Throughout, we are interested in the large $T$ limit where $z_i$ is small.
Taylor expanding Eq.~(\ref{eq_thermPot1b}) about $z_i=0$ to third order,
we find~\cite{LiuPRA2010}
\begin{align}
\label{eq_thermPotHighTa}
\Omega^{(2)} = & \; \Omega^{(1)}_1 + \; \Omega^{(1)}_2 + \Delta \Omega,
\end{align}
where
\begin{align}
\label{eq_singCompFGa}
\Omega_1^{(1)} & = -k_B T Q_{1,0} \sum_{n_1=1}^{\infty} b_{n_1,0} 
\; z_1^{n_1}, \\
\label{eq_singCompFGb}
\Omega_2^{(1)} & = -k_B T Q_{1,0} \sum_{n_2=1}^{\infty} b_{0,n_2} 
\; z_2^{n_2}
\end{align}
and
\begin{align}
\label{eq_thermPotHighTb}
\Delta \Omega = & - k_B T Q_{1,0} [\; b_{1,1} \; z_1 z_2 \; + \nonumber \\ 
& \; b_{1,2} \; z_1 z_2^2 \; + \; b_{2,1} \; z_1^2 z_2 \; + \; \cdots\; ].
\end{align}
Equations (\ref{eq_thermPotHighTa})-(\ref{eq_thermPotHighTb})
show that the thermodynamic potential of the two-component Fermi gas can
be written as a sum of the thermodynamic potentials $\Omega_1^{(1)}$
and $\Omega_2^{(1)}$ of components 1 and 2 and a correction term 
$\Delta\Omega$.
The thermodynamic potentials $\Omega_1^{(1)}$ and $\Omega_2^{(1)}$ of the
single-component species 1 and 2 are written in terms of the expansion
or virial coefficients $b_{n_1,0}$ and $b_{0,n_2}$ 
(see Appendix~\ref{sec_thermo1})
and the canonical partition function $Q_{1,0}$ of a single heavy 
particle of species 1.
The $b_{n_1,0}$ ($b_{0,n_2}$) with $n_1 > 1$ ($n_2>1$) 
arise from the fermionic nature of the atoms
and can be viewed as corrections to the Boltzmann gas~\cite{McQuarrie,HuangStatMech}.
The correction term $\Delta\Omega$ contains
the expansion or virial coefficients $b_{n_1,n_2}$, which account
for the interactions between distinguishable fermions~\cite{notation1,notation2},
\begin{align} 
\label{eq_v11}
b_{1,1} = & (Q_{1,1}-Q_{1,0}Q_{0,1})/Q_{1,0}, \\
\label{eq_v12}
b_{1,2} = & (Q_{1,2}-Q_{1,0}Q_{0,2} -b_{1,1}Q_{1,0}Q_{0,1})/Q_{1,0},
\end{align}
and
\begin{align}
\label{eq_v21}
b_{2,1} = & (Q_{2,1}-Q_{2,0}Q_{0,1} -b_{1,1}Q_{1,0}Q_{1,0})/Q_{1,0}.
\end{align}
In the non-interacting limit, i.e., for $a_s=0$, 
we label the partition functions by a superscript NI. In this case,
$Q_{n_1,n_2}^{\rm{NI}}$ reduces to $Q_{n_1,0}Q_{0,n_2}$,
since $H(n_1,n_2)$ is separable for $a_s=0$.
In this limit, 
$b_{1,1}$ vanishes since $Q^{\rm{NI}}_{1,1}$ equals $Q_{1,0}Q_{0,1}$.  
Consequently,
$b_{1,2}$ and $b_{2,1}$ as well as all higher order virial coefficients
$b_{n_1,n_2}$ with $n_1+n_2>3$ vanish; thus, $\Delta\Omega$ is zero
for vanishing $a_s$.

To determine the $b_{n_1,n_2}$, one must know three things: 
(i) all $b_{i,j}$ where $i \le n_1$ and $j<n_2$ or $i<n_1$ and $j \le n_2$;
(ii) all $Q_{i,0}$ and $Q_{0,j}$, where $i \le n_1$ and $j \le n_2$; and
(iii) the complete energy spectrum of the ($n_1,n_2$) system.
The $Q_{n_1,0}$ and $Q_{0,n_2}$ 
characterize the single-component Fermi gas and are
calculated in Appendix~\ref{sec_thermo1}.
In calculating the $Q_{n_1,n_2}$, it is convenient to express the total 
energy $E^{(n_1,n_2)}$ 
as a sum of the center of mass energy $E_{\rm{CM}}$ and the relative energy 
$E_{\rm{rel}}^{(n_1,n_2)}$\!,
$E^{(n_1,n_2)} = E_{\rm{CM}} + E^{(n_1,n_2)}_{\rm{rel}}\!\!$. 
In Eqs.~(\ref{eq_v11})-(\ref{eq_v21}), 
the center of mass contribution cancels
the $Q_{1,0}$ term in the denominator~\cite{LiuPRL2009}.

We now evaluate the $b_{n_1,n_2}$ at unitarity, i.e., for $1/a_s=0$.
The calculation of 
$b_{1,1}$ is straightforward~\cite{LiuPRL2009}.
The relative energies of the $(n_1,n_2)=(1,1)$ 
system in a harmonic trap can be solved 
exactly using zero-range $s$-wave interactions~\cite{Busch1998}.
The relative energies of the $(1,1)$ system at unitarity are
$E_{\rm{rel}}^{(1,1)} = (2 q+1/2) \hbar\omega$
for $l=0$ and $E_{\rm{rel}}^{(1,1)} = (2 q+l+3/2) \hbar\omega$ 
for $l=1,2,\ldots$, 
where $q$ is a non-negative integer.
As the relative energy spectrum for $l>0$ is identical to the single particle
energy spectrum, 
we have that $Q_{1,1}/Q_{1,0}$ is equal to $Q_{0,1}$ for states with $l>0$.
The remaining sum,
\begin{align}
\label{eq_b11a}
b_{1,1}=\sum_{q=0}^{\infty} \left( e^{-(2q+1/2)\tilde{\omega}} -
e^{-(2q+3/2)\tilde{\omega}} \right),
\end{align}
includes all $s$-wave states and can be calculated analytically~\cite{LiuPRL2009},
\begin{align}
\label{eq_b11b}
b_{1,1}= \frac{e^{\tilde{\omega}/2}}{1+e^{\tilde{\omega}}},
\end{align}
where $\tilde{\omega}$ denotes a dimensionless inverse temperature,
$\tilde{\omega}=\hbar\omega/(k_B T)$.
Equation (\ref{eq_b11b}) shows that $b_{1,1}$ is independent of 
the mass ratio.
This is a direct consequence of the fact that 
the relative two-body energy spectrum 
is independent of the mass ratio.
Furthermore, like the single-component virial coefficients $b_n$
(see Appendix~\ref{sec_thermo1}), $b_{1,1}$ is an even function in 
$\tilde{\omega}$.

To calculate $b_{1,2}$ and $b_{2,1}$, we need the complete relative energy
spectrum of the $(1,2)$ system (1 light atom and 2 heavy atoms) 
and the $(2,1)$ system (2 light atoms and 1 heavy atom), respectively.
The three-body energies are conveniently expressed in terms of the quantity
$\bar{\kappa}$,
\begin{align}
\label{eq_massratio}
\bar{\kappa} = 
\begin{cases}
1/\kappa & \text{for a majority of light species} \\
\kappa & \text{for a majority of heavy species}.
\end{cases}
\end{align}
At unitarity, the relative three-body energies for interspecies $s$-wave
zero-range interaction potentials can be written as
$E_{\rm{rel}}^{(n_1,n_2)}
=(2q+s_{l,\nu}+1)\hbar\omega$~\cite{WernerCastinPRA2006}, 
where $q=0,1,\ldots$,
and where the $s_{l,\nu}$ denote non-integer values.
The $s_{l,\nu}$ depend on the particle symmetry and $\bar{\kappa}$, and
are obtained by solving the five-dimensional hyperangular Schr\"odinger 
equation~\cite{WernerCastinPRL2006,OIKartavtsevJPB2007,RittenhousePRA2010}.
The hyperangular eigenvalues were first obtained for 
three identical bosons by Efimov~\cite{VNEfimov1970}, 
but have since been extended to 
any particle symmetry and mass ratio by both Efimov 
\cite{Efimov1973,EfimovLetter1972} and 
others
\cite{WernerCastinPRL2006,OIKartavtsevJPB2007,RittenhousePRA2010}.
Application of Ref. \cite{RittenhousePRA2010} to the two-component system with
$\bar{\kappa}$, relative angular momentum $l$, and 
parity $(-1)^l$ at unitarity yields
\begin{align}
\label{eq_trans}
-& \frac{\sqrt{4 \pi} \; \Gamma(l+3/2)}{\Gamma\!\left(\frac{1+l-s_{l,\nu}}{2}\right) 
 \Gamma\!\left(\frac{1+l+s_{l,\nu}}{2}\right)} =
\left(\frac{-\bar{\kappa}}{\bar{\kappa} + 1}\right)^l  \times \\
& _2F_1 \!\!
\left(1+\frac{l-s_{l,\nu}}{2},1+
\frac{l+s_{l,\nu}}{2},l+\tfrac{3}{2},
\frac{\bar{\kappa}^2}{(\bar{\kappa}+1)^2}\right)\!, \nonumber
\end{align}
where $\Gamma$ is the gamma function and 
$\!\;_2F_1$ is the hypergeometric function.
Equation~(\ref{eq_trans}) has a spurious 
root at $s_{0,\nu}=2$ for all $\bar{\kappa}$;
this root needs to be removed ``by hand".

\begin{figure}
\vspace*{+1.5cm}
\includegraphics[angle=0,width=70mm]{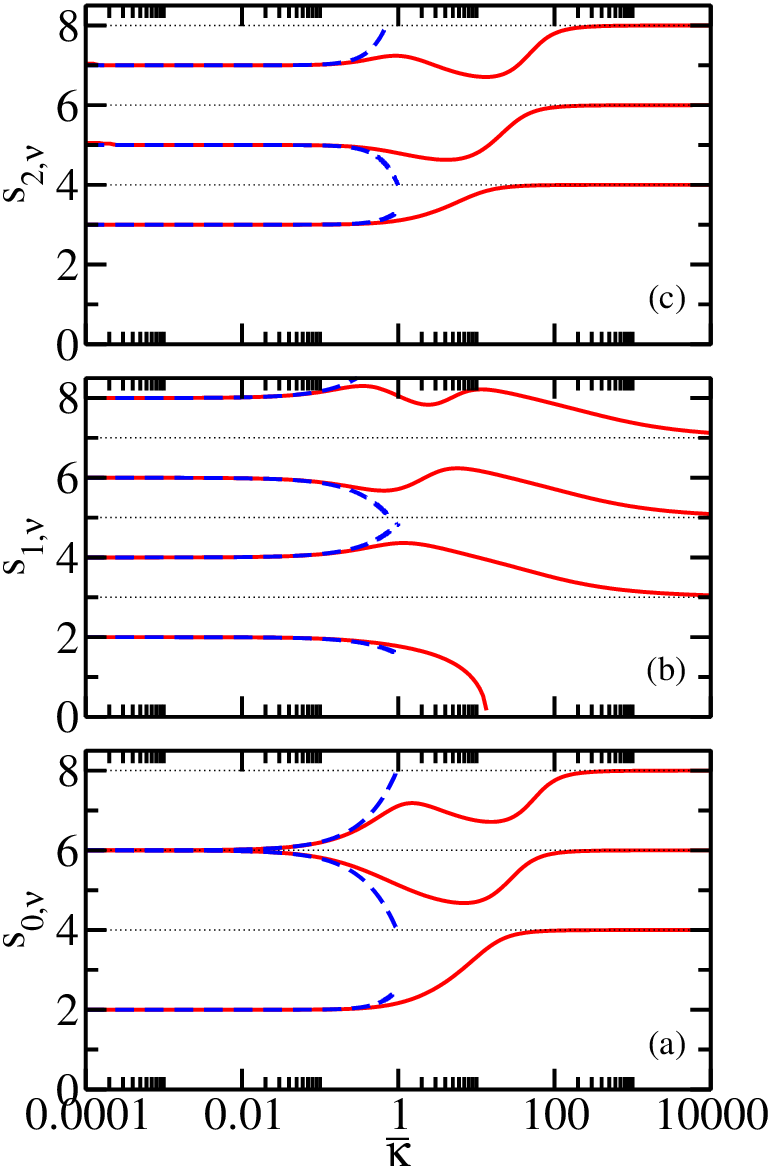}
\vspace*{0.1cm}
\caption{(Color online) Hyperangular eigenvalues $s_{l,\nu}$ 
as a function of $\bar{\kappa}$ for (a) $l=0$, (b) $l=1$, and 
(c) $l=2$ at unitarity. Note that $\bar{\kappa}$ is shown on a log scale.
Solid lines show the $s_{l,\nu}$ obtained by solving Eq.~(\ref{eq_trans}), 
dashed lines show the limiting 
expressions, Eqs.~(\ref{eq_transPerta})-(\ref{eq_transPertd}),
for $\bar{\kappa} \ll 1$, and thin dotted lines show the non-interacting
$s^{\rm{NI}}_{l,\nu}$.
}\label{fig_sln}
\end{figure}

Solid lines in Fig.~\ref{fig_sln} 
show the $s_{l,\nu}$ as a function of $\bar{\kappa}$ 
for the three lowest relative angular momenta, i.e., for $l=0$, 1, and 2.
The hyperangular quantum number $\nu$, $\nu=0,1,\ldots$,
counts the number of times that the slope of $s_{l,\nu}$
changes sign.
The thin dotted lines indicate the non-interacting limits, 
$s^{\rm{NI}}_{0,\nu}=2\nu+4$ for $l=0$ 
and $s^{\rm{NI}}_{l,\nu}=2\nu+l+2$ for $l>0$.
The dashed lines show an expansion of Eq.~(\ref{eq_trans}) for 
$\bar{\kappa} \ll 1$,
\begin{align}
\label{eq_transPerta}
s_{0,0} & \approx 2 + \frac{16}{3 \pi^2} \; \bar{\kappa}^2, \\
\label{eq_transPertb}
s_{0,2 j} & \approx 4 j + 2 + \frac{8}{\sqrt{3}\pi}\sqrt{j (j+1)} 
\; \bar{\kappa}, \\
\label{eq_transPertc}
s_{0,2j-1} & \approx 4 j + 2 - \frac{8}{\sqrt{3}\pi}\sqrt{j (j+1)} 
\; \bar{\kappa},
\end{align}
where $j=1$, 2, $\ldots$, and
\begin{align}
\label{eq_transPertd}
s_{l,\nu} & \approx 2 \nu + l + 1 + \frac{(-1)^{l+\nu} \; 
\Gamma(l+\nu+1)}{\sqrt{\pi} \; \Gamma(l+3/2) \; \Gamma(\nu+1)} 
\; \bar{\kappa}^l
\end{align}
for $l>0$.
For $l=0$, the lowest eigenvalue varies quadratically with 
$\bar{\kappa}$ in the
small $\bar{\kappa}$ regime. The higher lying $s_{l,\nu}$ values for 
$l=0$ appear in
pairs as $\bar{\kappa} \rightarrow 0$, with a splitting that is linear in 
$\bar{\kappa}$.
For $l>0$, in contrast, all eigenvalues are non-degenerate as 
$\bar{\kappa} \rightarrow 0$.  The lowest $l=1$ eigenvalue equals 2 
for $\bar{\kappa}=0$,
crosses 1 for $\bar{\kappa} \approx 8.62$, and becomes purely imaginary for 
$\bar{\kappa} \approx 13.61$.  The latter point indicates the onset of Efimov
physics~\cite{Petrov2003}.  For $8.62 \le \bar{\kappa} \le 13.61$, 
non-universal three-body states can exist~\cite{Petrov2003,WernerCastinPRA2006,NishidaSonTan2008,DBlumePRL2010,DBlumePRA2010}.
Similar behavior is found for the lowest $s_{l,\nu}$ values for higher odd
angular momenta.  The $s_{l,\nu}$ values for even $l$ are greater than 1 for
all $\bar{\kappa}$, indicating the absence of non-universal and 
Efimov physics in the even $l$ angular momentum channels.
For $l=1,3,5,\ldots$ [see Fig.~\ref{fig_sln}(b) for $l=1$], 
the $s_{l,\nu}$ with $\nu>0$ begin one unit below the non-interacting value in
the $\bar{\kappa}=0$ limit and decrease to the next lower non-interacting value
in the limit $\bar{\kappa}\rightarrow\infty$.
For $l=2,4,6,\ldots$ [see Fig.~\ref{fig_sln}(c) for $l=2$], 
the $s_{l,\nu}$ begin one unit below the non-interacting values in
the $\bar{\kappa}=0$ limit and approach the non-interacting values 
in the limit $\bar{\kappa} \rightarrow \infty$.  
In the limit $\bar{\kappa} \rightarrow \infty$, all states save those
that describe Efimov physics behave like non-interacting states.
The two masses are so heavy compared to the single light particle that
even infinitely strong interactions cannot mediate a lowering of the energy.

Now that we have the $s_{l,\nu}$, we can calculate 
$Q_{1,2}$ and $Q_{2,1}$, and thus $b_{1,2}$ and $b_{2,1}$.
We restrict ourselves to the regime where Efimov physics is absent (i.e., we
limit ourselves to \
$\bar{\kappa} \le 13.61$) and we assume that the three-body system 
behaves fully universal (i.e., we assume the absence of three-body resonances
in the regime $8.62 \le \bar{\kappa} \le 13.61$).  Under these assumptions,
the virial coefficient $b_{2,1}$, Eq.~(\ref{eq_v21}), can be written as
\begin{align}
\label{eq_b21a}
b_{2,1} = \sum_{l=0}^{\infty} b_{2,1}(l),
\end{align}
where
\begin{align}
\label{eq_b21}
b_{2,1}(l) = \sum_{\nu=0}^{\infty}\sum_{q=0}^{\infty}
(2l+1)\biggr[e^{-(2q+s_{l,\nu}+1)\tilde{\omega}}\nonumber \\
- e^{-(2q+s_{l,\nu}^{\rm{NI}}+1)\tilde{\omega}} 
- e^{-(2q+1/2+2\nu+l+3/2)\tilde{\omega}} \nonumber \\
+ e^{-(2q+3/2+2\nu+l+3/2)\tilde{\omega}}\biggr].
\end{align}
The first and second terms in the square bracket on the right hand side
of Eq.~\eqref{eq_b21}
arise from the $Q_{2,1}/Q_{1,0}$ and $Q_{2,0}$ terms, 
respectively, while the third
and fourth terms in the square bracket on the right hand side of 
Eq.~\eqref{eq_b21} arise from the $-b_{1,1}Q_{1,0}$ term.
For $l>0$, the second and fourth terms cancel, 
while the third term can be rewritten in terms of $s_{l,\nu}^{\rm{NI}}$.
Performing the sum over $q$ analytically, we find
\begin{align}
\label{eq_s12L}
b_{2,1}(l>0) = & \frac{e^{2\tilde{\omega}}}{e^{2\tilde{\omega}}-1} \times \\
& \biggl[(2l+1) \sum_{\nu=0}^{\infty}
\left(e^{-(s_{l,\nu}+1) \tilde{\omega}} - e^{-s^{\rm{NI}}_{l,\nu} 
\tilde{\omega}}\right)\biggr] \nonumber.
\end{align}
For $l=0$, the second and fourth terms do not completely cancel and
contribute, after the sum over $q$ is done analytically, a term proportional
to $\exp(-3\tilde{\omega})$.
Combining the first and third terms in the square bracket on the
right hand side of Eq.~\eqref{eq_b21}, we find
\begin{align}
\label{eq_s12}
b_{2,1}(l=0) = & \frac{e^{2\tilde{\omega}}}{e^{2\tilde{\omega}}-1} 
\biggl[e^{-3\tilde{\omega}} - e^{-2 \tilde{\omega}} +   \\ 
& \sum_{\nu=0}^{\infty}\left(e^{-(s_{0,\nu}+1)\tilde{\omega}}
- e^{-s^{\rm{NI}}_{0,\nu}\tilde{\omega}}\right)\biggr] \nonumber.
\end{align}
Since the mass ratio dependence only enters through the $s_{l,\nu}$, 
Eqs.~\eqref{eq_b21a}-\eqref{eq_s12} remain valid for $b_{1,2}$.
In practice, we calculate the first 10,000 $s_{l,\nu}$ values for each $l$, 
$l=0-50$, and determine the $b_{1,2}(l)$ and 
$b_{2,1}(l)$ using Eqs.~(\ref{eq_s12L}) and
(\ref{eq_s12}).
We find that we obtain better convergence if we employ the same cutoff on the
terms that involve $s_{l,\nu}$ and $s_{l,\nu}^{\rm{NI}}$ than if we perform the
sum over $\nu$ that involves the non-interacting 
$s_{l,\nu}^{\rm{NI}}$ values analytically.

\begin{figure}
\vspace*{+1.5cm}
\includegraphics[angle=0,width=70mm]{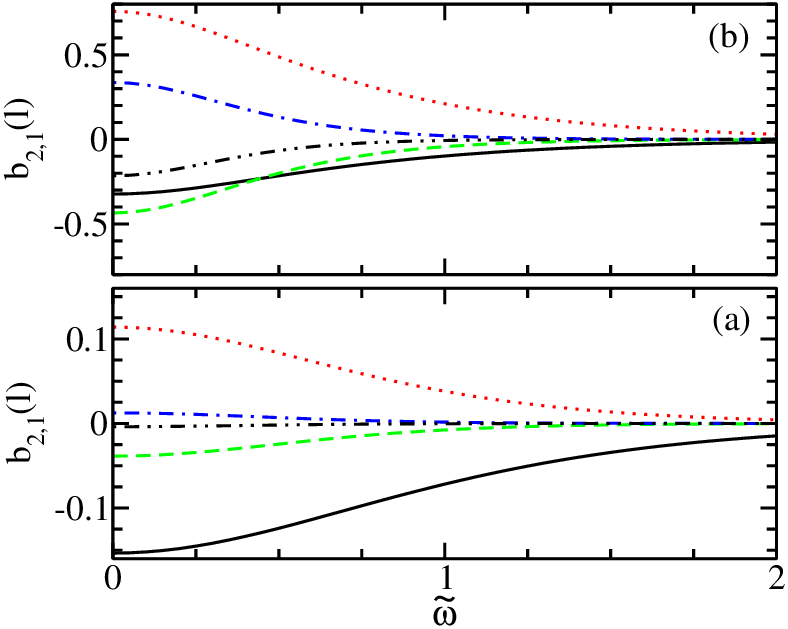}
\vspace*{0.1cm}
\caption{(Color online)
Angular momentum contributions $b_{2,1}(l)$ for $l=0-4$ for 
(a) $\bar{\kappa}=1$ and (b) $\bar{\kappa}=6.67$
as a function of $\tilde{\omega}$ at unitarity.  
Solid, dotted, dashed, dash-dotted, and dash-dot-dotted lines show 
$b_{2,1}(l)$ for $l$=0, 1, 2, 3, and 4, respectively.
}\label{fig_Lcontributions}
\end{figure}

Figure~\ref{fig_Lcontributions} 
shows the $b_{2,1}(l)$ for the first
five angular momenta for (a) $\bar{\kappa}=1$ and (b) $\bar{\kappa}=6.67$. 
The $b_{2,1}(l)$ are negative for even $l$ and positive for odd $l$ for all
$\tilde{\omega}$.
The leading contribution for any mass ratio
comes from the solution with $\nu=0$. For odd $l$, we have
$s_{l,0}+1<s_{l,0}^{\rm{NI}}$ [see Fig.~\ref{fig_sln}(b) for $l=1$], 
implying that the $b_{2,1}(l)$ 
are positive for all $\tilde{\omega}$ and $\bar{\kappa}$.
For even $l$, we have $s_{l,0}+1>s_{l,0}^{\rm{NI}}$ 
[see Fig.~\ref{fig_sln}(c) for $l=2$], implying that the $b_{2,1}(l)$
are negative.
For $l=0$, the $-\exp(-2\tilde{\omega})$ term gives the leading contribution, 
implying that $b_{2,1}(l)$ is negative for all $\tilde{\omega}$ 
and $\bar{\kappa}$.
A similar analysis can be conducted for the $b_{1,2}(l)$.
The alternating sign of the $b_{1,2}(l)$ and the $b_{2,1}(l)$ implies that
the full virial coefficients $b_{1,2}$ and $b_{2,1}$ converge like an 
alternating series as more $l$ terms are included.  We find that the 
convergence rate decreases as $\bar{\kappa}$ increases.

\begin{figure}
\vspace*{+1.5cm}
\includegraphics[angle=0,width=70mm]{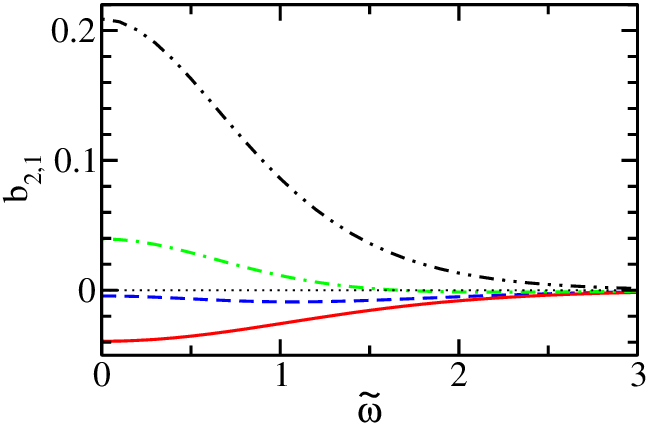}
\vspace*{0.1cm}
\caption{(Color online)
Solid, dashed, dash-dotted, and dash-dot-dotted
lines show the virial coefficient
$b_{2,1}$ as a function of $\tilde{\omega}$ for 
$\bar{\kappa}=2,3,4$ and 6.67. The thin dotted line at 0 is 
a guide to emphasize the sign change of $b_{2,1}$ for $\bar{\kappa}=4$ around 
$\tilde{\omega}\approx 1.67$.
}\label{fig_signChange234}
\end{figure}

Figure \ref{fig_signChange234}
shows the virial coefficients 
$b_{2,1}$ for $\bar{\kappa}=2,3,4$ and $6.67$ (from bottom to top) as a 
function of $\tilde{\omega}$.
The virial coefficient $b_{2,1}$ for $\bar{\kappa}=2$ and 3 is negative
for all $\tilde{\omega}$.
The virial coefficient for $\bar{\kappa}=4$ is negative for large 
$\tilde{\omega}$, changes sign at $\tilde{\omega} \approx 1.67$, 
and is positive for small $\tilde{\omega}$.
The virial coefficient for $\bar{\kappa}=6.67$ 
changes sign at a much lower temperature, i.e., at 
$\tilde{\omega} \approx 5.59$.
The sign change of $b_{2,1}$ can be attributed to the fact that the negative
$l=0$ contribution increases slower in magnitude than the positive $l=1$
contribution with increasing $\bar{\kappa}$ 
(see Fig.~\ref{fig_Lcontributions}), so that
the positive $l=1$ contribution dominates for sufficiently large 
$\bar{\kappa}$.
In the high temperature limit, $b_{2,1}$ first becomes positive for
$\bar{\kappa}\approx 3.11$.
In the low temperature regime, $b_{2,1}$ remains negative for 
$\bar{\kappa}<8.62$ and changes sign when $s_{1,0}=1$, i.e.,
the $\bar{\kappa}$ value beyond which three-body resonances can 
appear~\cite{Petrov2003,WernerCastinPRA2006,NishidaSonTan2008,DBlumePRL2010,DBlumePRA2010}.
We find that the virial coefficient $b_{1,2}$ is negative for all 
$\bar{\kappa}$ and $\tilde{\omega}$.
We note that a general framework for calculating $b_{1,2}$ and $b_{2,1}$ for
unequal-mass systems is given in Ref.~\cite{Leyronas2011};
however, this reference did not provide numerical values for
$b_{1,2}$ and $b_{2,1}$.
The fact that the third order virial coefficient $b_{2,1}$ changes sign
as $\bar{\kappa}$ increases suggests that 
the higher order virial coefficients might also change sign.
Throughout this paper, we restrict our analysis to systems with 
$n_1+n_2 \le 3$.

\begin{figure}
\vspace*{+1.5cm}
\includegraphics[angle=0,width=70mm]{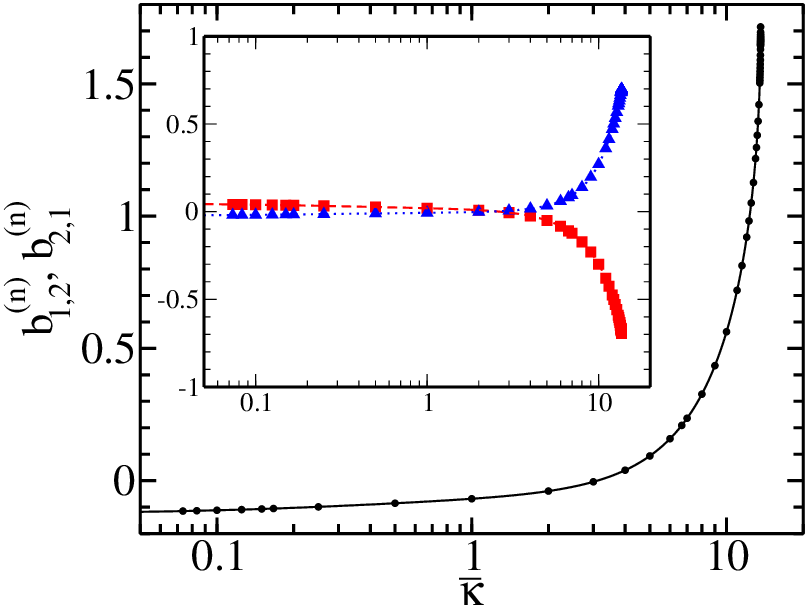}
\vspace*{0.1cm}
\caption{(Color online) 
Circles show the universal parts $b_{1,2}^{(0)}$ ($\bar{\kappa}<1$) 
and $b_{2,1}^{(0)}$ ($\bar{\kappa}>1$) of the virial coefficients
as a function of $\bar{\kappa}$.
The inset (the axes use the same labels as the main panel) shows
the first non-universal corrections $b_{1,2}^{(2)}$ and $b_{2,1}^{(2)}$
(squares) and
the second non-universal corrections $b_{1,2}^{(4)}$ and $b_{2,1}^{(4)}$
(triangles)
as a function of $\bar{\kappa}$.
The lines are guides to the eye.
}\label{fig_orders024}
\end{figure}

In the high temperature limit, we Taylor expand the $b_{n_1,n_2}$ 
($n_1+n_2=3$)~\cite{auxMaterial},
\begin{align}
\label{eq_bhighT}
b_{n_1,n_2} = b_{n_1,n_2}^{(0)} + b_{n_1,n_2}^{(2)}\tilde{\omega}^2 
+ b_{n_1,n_2}^{(4)}\tilde{\omega}^4 
+ \cdots.
\end{align}
To our calculated accuracy, we find that the odd powers vanish, i.e., the
virial coefficients $b_{1,2}$ and $b_{2,1}$ 
appear to be even functions in $\tilde{\omega}$, similar to
their single-component counterparts.
For the equal-mass system, the expression~\eqref{eq_bhighT} was considered
in Ref.~\cite{LiuPRL2009}.
The $b_{n_1,n_2}^{(0)}$ term is independent 
of $\tilde{\omega}$ and thus universal.
The higher order corrections $b^{(n>0)}_{1,2}$ and $b^{(n>0)}_{2,1}$ are
non-universal, since they depend on the harmonic trapping potential through 
$\tilde{\omega}$.
The supplementary material~\cite{auxMaterial} 
tabulates the $b_{1,2}^{(n)}$ and $b_{2,1}^{(n)}$,
$n=0,2,4$ and 6, as a function of $\bar{\kappa}$.
The $b_{1,2}^{(n)}$ and $b_{2,1}^{(n)}$ with $n=0,2,$ and 4 are shown in 
Fig.~\ref{fig_orders024} as a
function of $\bar{\kappa}$, where $\bar{\kappa}$ is shown on a log scale.
The universal virial coefficient $b_{2,1}^{(0)}$ has an infinite slope at
$\bar{\kappa}=13.61$ where Efimov physics sets in.  
For $\bar{\kappa} > 13.61$ (not shown), $b_{2,1}^{(0)}$ is expected to diverge
due to the contributions of the infinite sequence of Efimov trimers.
In the limit $\bar{\kappa}\rightarrow0$,
the $b_{1,2}(l)$ with $l>0$ vanish and we have
\begin{align}
\label{eq_b12w0limit}
b_{1,2}(\bar{\kappa}=0) = & \frac{-e^{2\tilde{\omega}}}
{(e^{\tilde{\omega}}+1)^2(e^{2 \tilde{\omega}} + 1)} \nonumber \\
\approx & -\frac{1}{8} + \frac{3}{32}\tilde{\omega}^2 
- \frac{3}{64}\tilde{\omega}^4 + \mathcal{O}(\tilde{\omega}^6).
\end{align}

\section{Thermodynamics of the trapped mass-imbalanced two-component fermi gas}
\label{sec_thermo2}
This section considers the thermodynamics of the harmonically trapped
two-component Fermi gas at unitarity as a function of $\kappa$. For the 
equal-mass case, we refer the reader to Refs.~\cite{LiuPRL2009,LiuPRA2010,HuNewJPhys2010,SalomonNature2010,TokyoScience2010}.
We measure our temperature $T$ in units of the semi-classical Fermi temperature
$T_F$ of a single-component Fermi gas with $N/2$ particles of mass $m_1$,
$T_F=[6(N/2)]^{1/3}\hbar \omega/k_B$~\cite{notation3}.
Correspondingly, our energy is measured in units of $E_F$, where $E_F=k_BT_F$.
Our starting point is the thermodynamic potential $\Omega^{(2)}$\!, 
Eq.~\eqref{eq_thermPotHighTa}, with $n_1 + n_2 \le 3$ (i.e., including
$b_{1,1}$, $b_{1,2}$, and $b_{2,1}$).

\begin{figure}
\vspace*{+1.5cm}
\includegraphics[angle=0,width=70mm]{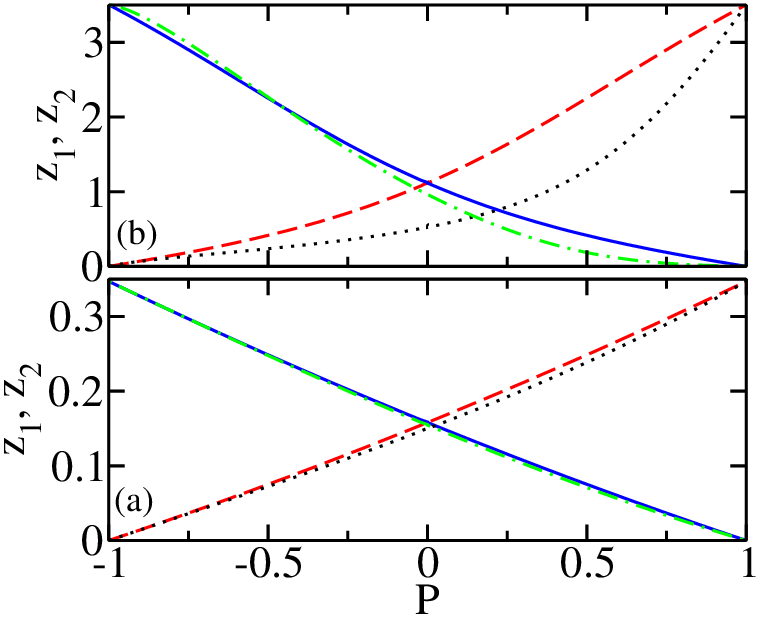}
\vspace*{0.1cm}
\caption{(Color online) Trapped system at unitarity.
Solid and dashed lines (dash-dotted and dotted lines)
show the fugacities $z_1$ and $z_2$, respectively, for $\kappa=1$ 
($\kappa=13$) as a function of the polarization $P$ for 
$N=5\times 10^7$ and
(a) $T/T_F=1$ and
(b) $T/T_F=1/2$.
}\label{fig_fugacities}
\end{figure}

The two coupled number equations, derived from 
$N_i=-\partial \Omega^{(2)} / \partial \mu_i$ ($i=1$ and 2), are
\begin{align}
\label{eq_N1}
\frac{N_1}{Q_{1,0}} = f_1(z_1)
+  b_{1,1}z_1z_2 + 2 b_{2,1}z_1^2z_2 + b_{1,2}z_1z_2^2
\end{align}
and
\begin{align}
\label{eq_N2}
\frac{N_2}{Q_{1,0}} = f_2(z_2) 
+  b_{1,1}z_1z_2 + b_{2,1}z_1^2z_2 + 2 b_{1,2}z_1z_2^2.
\end{align}
Throughout, we use the full temperature dependent expressions for
the virial coefficients, and
\begin{align}
\label{eq_fz1}
f_1(z_1) = \sum_{n_1=0}^{\infty} n_1 b_{n_1,0} z_1^{n_1}
\end{align}
and
\begin{align}
\label{eq_fz2}
f_2(z_2) = \sum_{n_2=0}^{\infty} n_2 b_{0,n_2} z_2^{n_2}.
\end{align}
Equations~\eqref{eq_fz1} and~\eqref{eq_fz2} can be employed if
$z_i < e^{3\tilde{\omega}/2}$, $i=1,2$.
For larger $z_i$,
useful parameterizations of Eqs.~\eqref{eq_fz1} and~\eqref{eq_fz2} are
\begin{align}
\label{eq_fz}
f_i(z_i) & \approx -\mbox{Li}_3(-z_i)  \\
& + \tfrac{1}{8} \tilde{\omega}^2 
\left[-\mbox{Ln}(1+z_i)-\mbox{Li}_3(-z_i) \right] \nonumber \\
& + \tfrac{1}{1920} \tilde{\omega}^4 \left[ \frac{17 z_i}{(1+z_i)^2} - 
30 \mbox{Ln}(1+z_i) - 13 \mbox{Li}_3(-z_i) \right] \nonumber \\
& + \mathcal{O}(\tilde{\omega}^6), \nonumber
\end{align}
where Ln is the natural logarithm function and Li is the polylogarithm 
function.
For the temperatures considered in this paper, terms of order 
$\tilde{\omega}^6$ and higher are negligible.
Equations~\eqref{eq_fz1} and~\eqref{eq_fz2} treat the non-interacting
``reference pieces'' $f_1(z_1)$ and $f_2(z_2)$ as an infinite series
in the fugacities and the interacting pieces at third order in the fugacities.
Alternatively, one may consider truncating the non-interacting and interacting
pieces at the same order in the fugacities.
We have opted for the former approach for two reasons:
(i) In the limit that $P=1$ or $-1$, we recover the exact results of the
single-component Fermi gas.
(ii) In the regime where the $z_i$ are small, i.e., in the regime where the
virial equation of state up to third order is expected to provide reliable
results, we find fairly small differences 
[e.g., $<0.06\%$ in Fig.~\ref{fig_fugacities}(a)]
between the approaches that
treat the non-interacting piece at infinite order and third order in the
fugacities.
For a given temperature $T$, polarization $P$, and total number of particles
$N$, 
we solve Eqs.~\eqref{eq_N1} and \eqref{eq_N2} 
self-consistently for $z_1$ and $z_2$.

Figure~\ref{fig_fugacities}
shows the fugacities $z_1$ and $z_2$ at unitarity
as a function of the polarization for
(a) $T/T_F=1$ and (b) $T/T_F=1/2$ for $\kappa=1$ (solid and dashed lines)
and $\kappa=13$ (dash-dotted and dotted lines).
The fugacities for $T/T_F=1$ are smaller than 
0.35 for both mass ratios considered,
suggesting that the virial equation of state can 
likely be trusted for $T/T_F \ge 1$.
For $T/T_F=1/2$, in contrast, the fugacities are as large as 3.5, 
suggesting that the description is at best qualitatively correct.
On the scale shown, the fugacities for $T/T_F=1$ and $\kappa=1$ are nearly
indistinguishable from those at the same temperature but higher mass ratio.
This makes sense intuitively, since the fugacities are small 
in the high temperature regime,
thereby suppressing higher order terms in the virial expansion.  
At lower temperatures [see Fig.~\ref{fig_fugacities}(b)], the higher order
virial coefficients become more important, as evidenced by the fact that the
fugacities show a notable dependence on the mass ratio.

\begin{figure}
\vspace*{+1.5cm}
\includegraphics[angle=0,width=70mm]{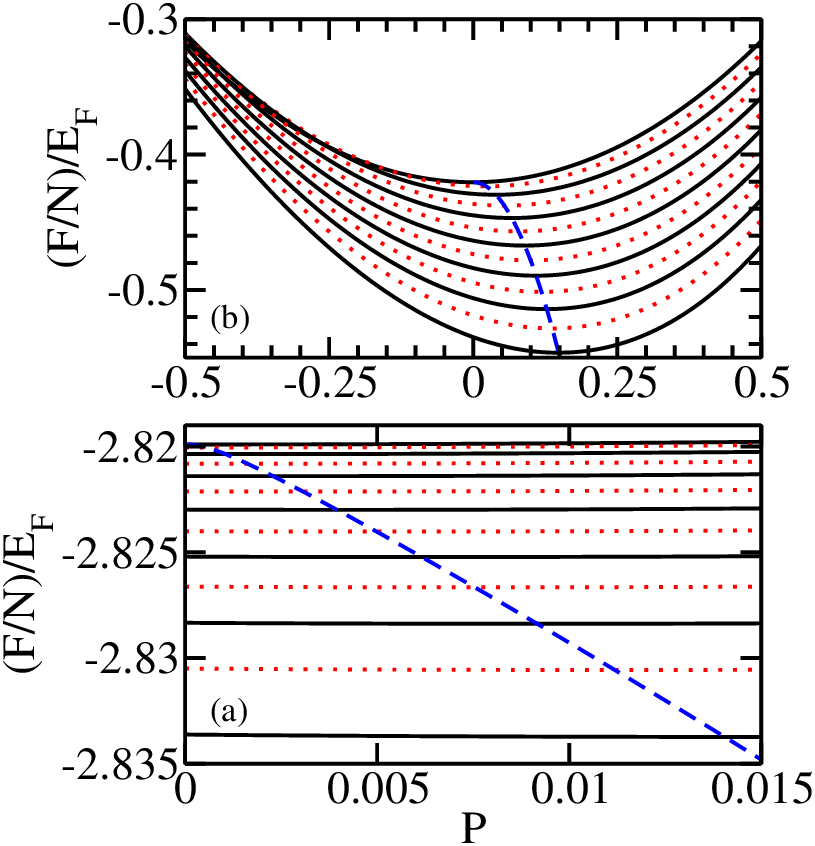}
\vspace*{0.1cm}
\caption{(Color online) Free energy per particle $F/N$ ($N=5\times 10^7$)
as a function of the polarization $P$ for
(a) $T/T_F=1$ and
(b) $T/T_F=1/2$
for the trapped system at unitarity.
Alternating solid and dotted curves, from top to bottom, 
show the free energy for $\kappa=1-13$ in units of 1.
The dashed lines connect the $(F_{eq},P_{eq})$ points for different $\kappa$.
}\label{fig_freeEnergy1}
\end{figure}

In the following, we use the virial equation of state to 
determine the free energy $F$, 
$F = \Omega^{(2)} + \mu_1 N_1 + \mu_2 N_2$, the entropy $S$,
$S=-\partial \Omega^{(2)}/\partial T$ 
(calculated for fixed $\omega$, $\mu_1$ and $\mu_2$),
and the energy $U$, $U=F+TS$, as a function of the 
mass ratio $\kappa$, the polarization $P$, and the temperature $T$
at unitarity.
Solid and dotted lines in Fig.~\ref{fig_freeEnergy1}
show the free energy per particle $F/N$ as a function of the polarization for
(a) $T/T_F=1$ and (b) $T/T_F=1/2$.
In both panels, the alternating solid and dotted lines, from top to bottom,
correspond to mass ratios $\kappa=1,2,\ldots,13$.
For the equal mass system, the minimum of the free energy occurs for $P=0$,
i.e., for equal numbers of spin-up and spin-down fermions.
As $\kappa$ increases, the minimum of the free energy moves to positive
polarizations, i.e., to a majority of heavy particles.  We refer to the
$(F,P)$ values at which the free energy is minimized as $(F_{eq},P_{eq})$.
Dashed lines in Figs.~\ref{fig_freeEnergy1}(a) and \ref{fig_freeEnergy1}(b)
connect the $(F_{eq},P_{eq})$ values for different $\kappa$.
For both temperatures, $P_{eq}$ increases with increasing $\kappa$,
which can be understood by realizing that the three-body system
consisting of two heavy atoms and one light atom has a lower energy than
the three-body system consisting of one heavy atom and two light atoms.
Moreover, for a given $\kappa$, $P_{eq}$ increases with decreasing temperature.
This trend makes sense, since we expect interaction 
effects to become more important as the temperature decreases.

To obtain an analytical expression for $P_{eq}$, we take the derivative of
$F$ with respect to $P$ at fixed temperature and fixed number of particles.
We find that the extremum of $P$ occurs when the two fugacities are equal
(i.e., $z_1=z_2=z_{eq}$).
This restriction allows for a straightfoward calculation of $z_{eq}$, namely
\begin{align}
\label{eq_zeq}
\frac{N}{Q_{1,0}} = f_1(z_{eq})+f_2(z_{eq})+2 b_{1,1}z_{eq}^2 
+ 3(b_{2,1}+b_{1,2})z_{eq}^3
\end{align}
can be solved self-consistently for fixed $N$ and $T$.
Up to third order in $z_{eq}$, $P_{eq}$ can be written as
\begin{align}
\label{eq_Feq}
P_{eq} = \frac{N}{Q_{1,0}} (b_{2,1}-b_{1,2}) z_{eq}^3.
\end{align}
$P_{eq}$ is equal to
zero for all $\kappa$ if the virial equation of state terminates at the 
second order in the fugacity.  At the third order, we find $P_{eq}=0$ for
$\kappa=1$, since $b_{1,2}$ equals $b_{2,1}$ in this case.
For $\kappa>1$, however, $P_{eq}$ is positive since $b_{2,1}>b_{1,2}$ for all
$\kappa>1$. 
Interestingly, the sign change of $b_{2,1}$ in the high temperature
regime for $\kappa \approx 3.11$ does not lead to a discontinuity or a sign
change of $P_{eq}$ since $b_{1,2}$ is finite and negative for all $\kappa$.
For small deviations from $\kappa=1$, we find that $P_{eq}$ and $F_{eq}$ change
linearly and quadratically, respectively, as a function of $\kappa$ for
fixed temperature and fixed number of particles. We find that
$\partial^2 F/\partial P^2$, calculated for fixed $T$ and $N$, is greater
than zero for all polarizations and mass ratios considered, which implies
that the system is stable at this level of approximation.

\begin{figure}
\vspace*{+1.5cm}
\includegraphics[angle=0,width=70mm]{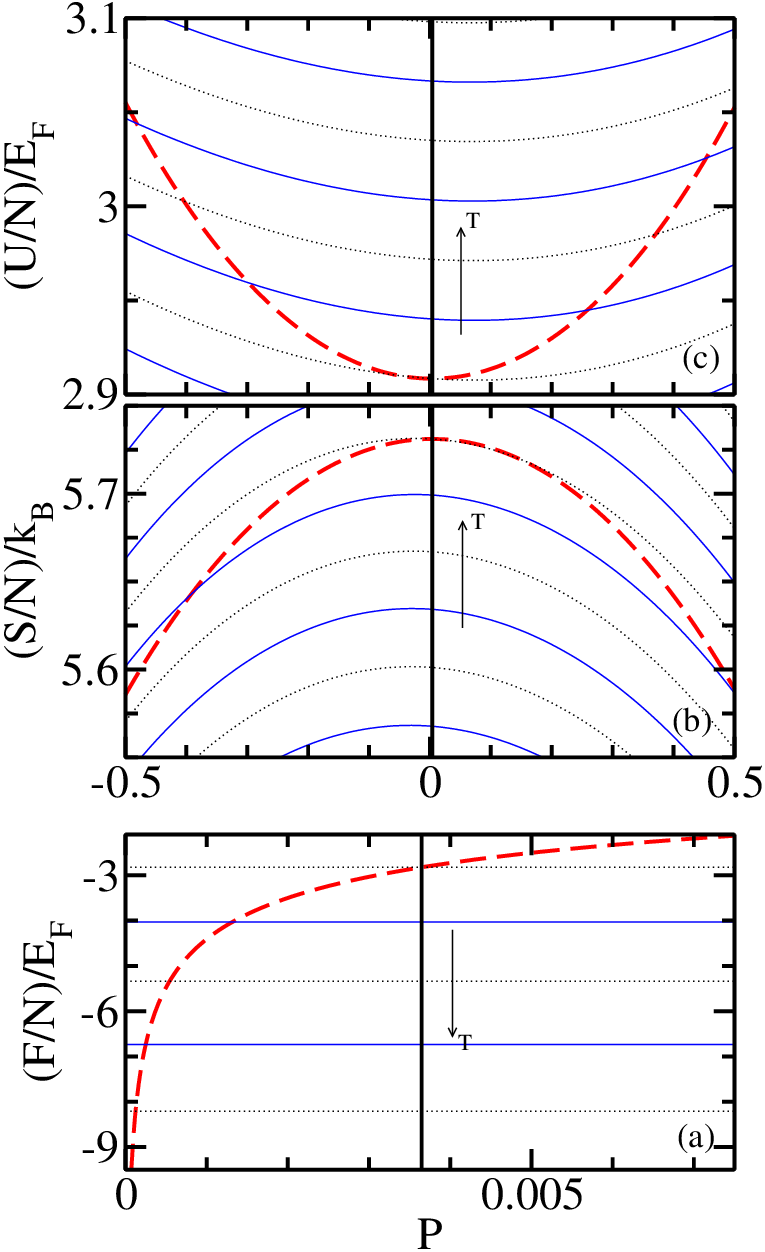}
\vspace*{0.1cm}
\caption{(Color online) Thermodynamic observables $F/N$, $S/N$, 
and $U/N$ as a function
of the polarization $P$ for $\kappa=6.67$ and $N=5\times 10^7$
for the trapped system at unitarity.
Alternating dotted and solid lines show
(a) the free energy per particle $F/N$ 
for $T/T_F=1,1.2,\ldots,1.8$ (from top to bottom),
(b) the entropy per particle $S/N$
for $T/T_F=0.95,0.96,\ldots,1.04$ (from bottom at $P\approx 0$ 
to top at $P \approx 0.5$), and 
(c) the energy per particle $U/N$
for $T/T_F=0.99,1,\ldots ,1.06$ (from bottom at $P \approx 0.5$ 
to top at $P\approx0.1$).
In panel (a), the dashed line connects
$(F_{eq},P_{eq})$ values for different $T$.
In panels (b) and (c), the dashed lines show $S/N$ for $U/N=2.91E_F$ and
$U/N$ for $S/N=5.73k_B$, respectively.
In all three panels,
the vertical lines indicate the value of $P_{eq}$ for $T/T_F=1$.
}\label{fig_thermoCombined}
\end{figure}

To discuss the dependence of the entropy $S$ and the energy $U$ on the
temperature and the polarization, we focus on the mass ratio $\kappa=6.67$,
i.e., on the experimentally realizable K-Li system at unitarity~\cite{EWillePRL2008,MTaglieberPRL2008,FMSpiegelhalderPRA2010,LCostaPRL2010,ARidingerEPJD2011,DNaikEPJD2011,ATrenkwalderPRL2011}.
Thin dotted and solid lines in Fig.~\ref{fig_thermoCombined}(a)
show the free energy per particle as a function of the polarization for
$T/T_F=1,1.2,\ldots,1.8$ (from top to bottom).
The dashed line connects the $(F_{eq},P_{eq})$ values for different $T$
(but fixed $\kappa$ and $N$).
The vertical solid line marks the value $P_{eq}$ for $T/T_F=1$, 
where $F/N=-2.82E_F$, $S/N=5.73k_B$, and $U/N=2.91E_F$.
Thin solid and dotted lines in
Fig.~\ref{fig_thermoCombined}(b) show the entropy per particle $S/N$ as
a function of the polarization for $T/T_F=0.95,0.96,\ldots,1.04$ (from bottom
at $P\approx 0$ to top at $P \approx 0.5$).
Similarly, thin solid and dotted lines in Fig.~\ref{fig_thermoCombined}(c)
show the energy per particle $U/N$ as a function of the polarization for
$T/T_F=0.99,1,\ldots ,1.06$ (from bottom at $P \approx 0.5$ 
to top at $P\approx0.1$).
The dashed lines in Figs.~\ref{fig_thermoCombined}(b) 
and~\ref{fig_thermoCombined}(c) connect points of constant $U/N=2.91E_F$
and $S/N=5.73k_B$, respectively (i.e., the values of $U/N$ and $S/N$
at $P_{eq}$ for $T/T_F=1$).
As expected, the entropy for a fixed energy and the energy for a fixed 
entropy are concave and convex functions of the polarization, respectively.
It can be seen that the entropy per particle $S/N$ is maximized at $P_{eq}$
while the energy per particle $U/N$ is minimized at $P_{eq}$.
The fact that $-\partial^2 S/\partial P^2$ ($\partial^2 U / \partial P^2$) is
greater than zero for fixed $U$ and $N$ (for fixed $S$ and $N$) is
consistent with the fact that the system is stable as concluded earlier from
the fact that $\partial^2 F / \partial P^2$ is greater than zero for fixed
$T$ and $N$.  Lastly, we note that the maximum of the isotherm $S/N$ 
moves to smaller $P$ with decreasing $T$ and that the minimum of 
the isotherm $U/N$ moves to larger $P$ with decreasing $T$.

\begin{figure}
\vspace*{+1.5cm}
\includegraphics[angle=0,width=70mm]{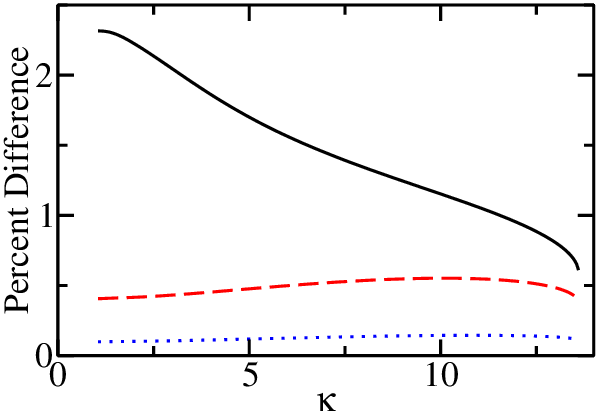}
\vspace*{0.1cm}
\caption{(Color online) Percent difference between $P_{eq}$ for 
$N=5 \times 10^7$ and $P_{eq}$ for $N=500$ as a function of 
the mass ratio $\kappa$ for the trapped system at unitarity.
Solid, dashed and dotted lines show the percent difference for $T/T_F=1/2,1$ 
and 2, respectively.
}\label{fig_percentDiff}
\end{figure}

The thermodynamic quantities shown in 
Figs.~\ref{fig_fugacities}-\ref{fig_thermoCombined} 
have been obtained for $N=5 \times 10^7$ particles.
For fixed $T/T_F$, the actual temperature $T$ decreases with decreasing $N$, 
since
$T_F$ decreases with decreasing $N$.  Correspondingly, $\tilde{\omega}$
increases with decreasing $N$ (again, assuming fixed $T/T_F$).  
Since the leading
order non-universal corrections scale with $\tilde{\omega}^2$ 
[see Eqs.~\eqref{eq_bhighT} and~\eqref{eq_fz}],
the non-universal corrections are expected to be negligible for large $N$,
but not for very small $N$.  
Figure~\ref{fig_percentDiff}
shows the percentage difference
between $P_{eq}$ for $N=5\times 10^7$ and $P_{eq}$ for $N=500$ as a function
of $\kappa$ for $T/T_F=2,1$ and 1/2.  The non-universal corrections are 
largest,
i.e., on the order of 2\%, for $\kappa=1$ and $T/T_F=1/2$ 
[solid line in Fig.~\ref{fig_percentDiff}].
For larger $N$, the non-universal corrections are even smaller, implying
that Figs.~\ref{fig_fugacities}-\ref{fig_thermoCombined} show essentially
universal, $N$-independent thermodynamic quantities of the trapped
two-component Fermi system with equal trapping frequencies.

We conclude this section by noting that $P_{eq}$ is quite small for the 
K-Li system
down to $T/T_F=1$, but then changes more rapidly as the temperature is lowered
further.
The dependence of $P_{eq}$ on the mass ratio $\kappa$ may be enhanced by
switching to a trap geometry for which the single-particle 
energies are dependent
on the mass $m_1$ or $m_2$.  One such system is, as detailed in the 
next section,
the homogeneous system, which favors much larger polarizations than the trapped
system for comparable $T/T_F$.

\section{Thermodynamics of the universal homogeneous mass-imbalanced two-component fermi gas}
\label{sec_thermo3}

\subsection{Local Density Approximation}
This section shows how the homogeneous system can be related to
the harmonically trapped system via the local density approximation~\cite{LiuPRL2009,CMenottiPRL2002}.
In the high temperature limit, the thermodynamic potential 
$\Omega^{(2)}_{\rm{hom}}$ of the homogeneous system can be written as
\begin{align}
\label{eq_homHighT}
\Omega^{(2)}_{\rm{hom}} = \Omega^{(1)}_{1,\rm{hom}} 
+ \Omega^{(1)}_{2,\rm{hom}} 
+ \Delta\Omega_{\rm{hom}},
\end{align}
where 
\begin{align}
\label{eq_homSingCompFGa}
\Omega_{1,\rm{hom}}^{(1)} & = -k_B T \frac{V}{\lambda_{m_1}^3} 
\sum_{n_1=1}^{\infty} b_{n_1,0,\rm{hom}}  \; z_1^{n_1}, \\
\label{eq_homSingCompFGb}
\Omega_{2,\rm{hom}}^{(1)} & = -k_B T \frac{V}{\lambda_{m_1}^3} 
\sum_{n_2=1}^{\infty} b_{0,n_2,\rm{hom}}  \; z_2^{n_2},
\end{align}
and
\begin{align}
\label{eq_homThermPotHighTb}
\Delta \Omega_{\rm{hom}} = & - k_B T  \frac{V}{\lambda_{m_1}^3} 
\sum_{n_1=1}^{\infty} \sum_{n_2=1}^{\infty} b_{n_1,n_2,\rm{hom}}  
z_1^{n_1} z_2^{n_2}.
\end{align}
Here, the $b_{n_1,n_2,\rm{hom}} $ denote 
the virial coefficients of the homogeneous system, 
and $z_1$ and $z_2$ are the fugacities of the homogeneous system.
The virial coefficients $b_{n_1,n_2,\rm{hom}}$ are defined by 
Eqs.~\eqref{eq_v11}-\eqref{eq_v21}
with the $Q_{n_1,n_2}$ interpreted as the partition functions of the
homogeneous systems.
Equations~\eqref{eq_homHighT}-\eqref{eq_homThermPotHighTb} are analogous to
the high temperature limits of 
Eqs.~\eqref{eq_thermPotHighTa}-\eqref{eq_thermPotHighTb}.
In Eqs.~\eqref{eq_homHighT}-\eqref{eq_homThermPotHighTb}, we used that the
high temperature limit of $Q_{1,0,\rm{hom}}$ 
is given by $V/\lambda_{m_1}^3$~\cite{HuangStatMech},
where $V$ is the volume of the system and where 
$\lambda_{m_1}$ is the thermal de Broglie wavelength of a particle of the
heavy species,
\begin{align}
\label{eq_deBroglieWL}
\lambda_{m_1} = \sqrt{\frac{2 \pi \hbar^2}{m_1 k_B T}}.
\end{align}

In the local density approximation~\cite{LiuPRL2009,CMenottiPRL2002}, 
the thermodynamic potential $\Omega^{(2)}$
of the trapped system is given by a weighted average of
$\Omega^{(2)}_{\rm{hom}}(\vec{r})$ over all space,
\begin{align}\label{eq_N1trap}
\Omega^{(2)} = \frac{\int \Omega^{(2)}_{\rm{hom}}(\vec{r}) 
d^3 \vec{r}}{\int d^3 \vec{r}},
\end{align}
i.e., we assume that the thermodynamic potential at each location $\vec{r}$
in the trap can be approximated by the corresponding bulk value.
This implies that the $z_i$ in 
Eqs.~\eqref{eq_homSingCompFGa}-\eqref{eq_homThermPotHighTb} 
are replaced by $z_i(\vec{r})$,
$z_i(\vec{r})=\exp[\mu_{i,loc}(\vec{r})/(k_B T)]$,
where the local chemical potentials $\mu_{i,loc}(\vec{r})$ are defined as
\begin{align}\label{eq_chemPotLoc}
\mu_{i,loc}(\vec{r}) = \mu_i - \tfrac{1}{2}m_i\omega^2\vec{r}^2.
\end{align}

The integrations on the right hand side of Eq.~\eqref{eq_N1trap} are
straightforward.
We first perform the integration involving $\Delta\Omega_{\rm{hom}}(\vec{r})$.
The integration in the denominator introduces a factor of $V$, 
which cancels with the factor of $V$ in the numerator.
This leaves, after performing the angular integrals,
\begin{align}
\label{eq_locDensityInt}
\Delta\Omega = & -4 \pi \frac{k_B T}{\lambda_{m_1}^3} 
\sum_{n_1=1}^{\infty} \sum_{n_2=1}^{\infty}
b_{n_1,n_2,\rm{hom}} z_1^{n_1} z_2^{n_2} \times \\
&\int_0^{\infty}  \!\!\! 
\exp{[-\tfrac{1}{2}(m_1 n_1 + m_2 n_2)\omega^2 r^2/(k_B T)]r^2 dr}. \nonumber
\end{align}
Performing the remaining integration and 
expressing the result in terms of the mass ratio $\kappa$, we find
\begin{align}
\label{eq_O1trapA}
\Delta\Omega = & -k_B T \left(\frac{k_B T}{\hbar \omega}\right)^3 \times 
\nonumber \\
& \sum_{n_1=1}^{\infty} \sum_{n_2=1}^{\infty} 
\frac{b_{n_1,n_2,\rm{hom}} }{(n_1 + n_2 /\kappa)^{3/2}} z_1^{n_1} z_2^{n_2}.
\end{align}
The prefactor $[k_B T/(\hbar\omega)]^3$ coincides with the high $T$ limit of
$Q_{1,0}$ for the trapped system.
Comparison of Eqs.~\eqref{eq_thermPotHighTb} and~\eqref{eq_O1trapA}
allows us to relate the universal virial
coefficients $b_{n_1,n_2}^{(0)}$ of the trapped system 
to the virial coefficients $b_{n_1,n_2,\rm{hom}}$ of the
homogeneous system,
\begin{align}
\label{eq_vcHomToTrap}
b_{n_1,n_2,\rm{hom}}  = (n_1 + n_2 /\kappa)^{3/2} b_{n_1,n_2}^{(0)}.
\end{align}
Considering the integrations over $\Omega^{(1)}_{1,\rm{hom}}(\vec{r})$ and
$\Omega^{(1)}_{2,\rm{hom}}(\vec{r})$, one finds analogous relationships between
$b_{n_1,0,\rm{hom}} $ and $b_{n_1,0}^{(0)}$, and between
$b_{0,n_2,\rm{hom}} $ and $b_{0,n_2}^{(0)}$.

\subsection{Universal thermodynamics of the homogeneous system}
\label{sec_thermoHom}
This section considers the universal, $N$-independent thermodynamics of the
homogeneous two-component
Fermi gas at unitarity as a function of $\kappa$.
We measure our temperature $T$ in units of the semi-classical Fermi temperature
$T_F$ of a single-component Fermi gas with $N/2$ particles of mass $m_1$,
$k_B T_F=[6 \pi^2 (N/2)/V]^{2/3} \hbar^2/(2 m_1)$~\cite{HuangStatMech}.
Notice that the Fermi temperature scales inversely with the mass $m_1$.
This implies that the Fermi temperature of $N/2$ light particles with mass 
$m_2$ is $\kappa$ times larger than the Fermi temperature of $N/2$ heavy
particles.  The different Fermi temperatures of the single-component heavy and
light particles are, of course, a direct consequence of the mismatching Fermi
surfaces of unequal-mass Fermi gases or, equivalently, a direct consequence of
the fact that the single-particle energies of a mass $m_i$ particle in a box
with periodic boundary conditions scale with $1/m_i$.
The high temperature virial equation of state of the homogeneous system can be
applied down to $T/T_F$ about $1$ for $\kappa=1$.  For unequal masses, the
applicability regime, using the heavy mass to define $T_F$ (see above),
is limited to $T/T_F \gtrsim \kappa$.

\begin{figure}
\vspace*{+1.5cm}
\includegraphics[angle=0,width=80mm]{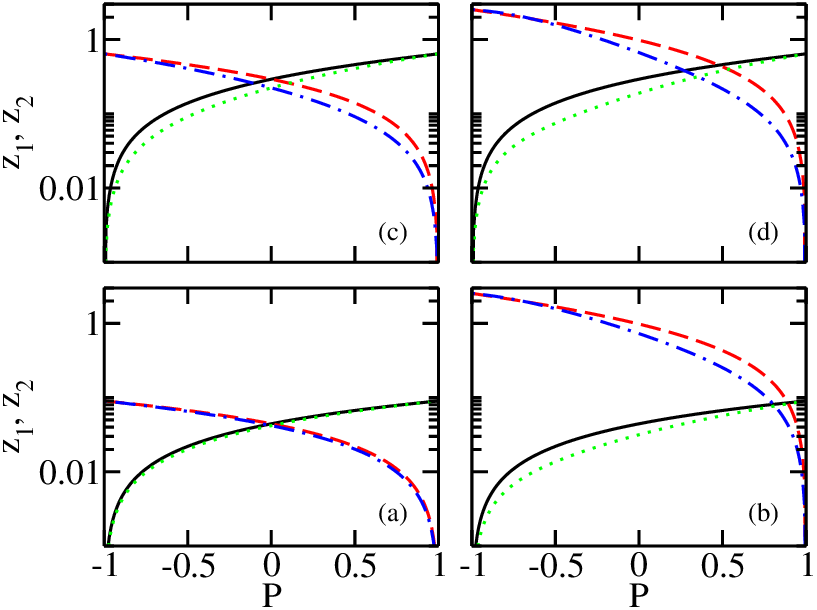}
\vspace*{0.1cm}
\caption{(Color online) Fugacities $z_1$ and $z_2$ of the homogeneous
system as a function of the polarization $P$ for
(a) $\kappa=1$ and $T/T_F=6.67$,
(b) $\kappa=6.67$ and $T/T_F=6.67$,
(c) $\kappa=1$ and $T/T_F=2$, and
(d) $\kappa=2$ and $T/T_F=2$.
Dotted and dash-dotted lines show $z_1$ and $z_2$, respectively,
for the unitary system obtained by solving Eqs.~\eqref{eq_homN1} 
and~\eqref{eq_homN2} self-consistently.
For comparison, solid and dashed lines show 
$z_1$ and $z_2$, respectively, for the non-interacting system.
}\label{fig_homFugacities}
\end{figure}

The coupled number equations of the
homogeneous system, up to third order, read
\begin{align}
\label{eq_homN1}
\frac{N_1 \lambda_{m_1}^3}{V} = & -\mbox{Li}_{3/2}(-z_1) + b_{1,1,\rm{hom}}  
z_1 z_2 \nonumber \\
& + 2 b_{2,1,\rm{hom}}  z_1^2 z_2 + b_{1,2,\rm{hom}}  z_1 z_2^2
\end{align}
and
\begin{align}
\label{eq_homN2}
\kappa^{3/2} & \frac{N_2 \lambda_{m_1}^3}{V} = -\mbox{Li}_{3/2}(-z_2) 
+ \kappa^{3/2} b_{1,1,\rm{hom}}  z_1 z_2 \nonumber \\
& + \kappa^{3/2} b_{2,1,\rm{hom}}  z_1^2 z_2 + 2 
\kappa^{3/2} b_{1,2,\rm{hom}}  z_1 z_2^2.
\end{align}
Dotted and dash-dotted lines in Fig.~\ref{fig_homFugacities}
show $z_1$ and $z_2$, respectively, as a function of the polarization $P$
for the unitary system obtained by solving Eqs.~\eqref{eq_homN1} 
and~\eqref{eq_homN2} self-consistently for fixed $\kappa$ and $T$.
The panels show the fugacities for 
(a) $\kappa=1$ and $T/T_F=6.67$,
(b) $\kappa=6.67$ and $T/T_F=6.67$,
(c) $\kappa=1$ and $T/T_F=2$, and
(d) $\kappa=2$ and $T/T_F=2$.
For the equal mass case, Eqs.~\eqref{eq_homN1} and~\eqref{eq_homN2}
are symmetric and thus $z_1$ and $z_2$ are symmetric about $P=0$ for
any temperature [see~Figs.~\ref{fig_homFugacities}(a) 
and~\ref{fig_homFugacities}(c)].
For unequal masses, in contrast, Eqs.~\eqref{eq_homN1} and~\eqref{eq_homN2}
are not symmetric and the crossing of the fugacities $z_1$ and $z_2$ is shifted
to positive polarizations [see Figs.~\ref{fig_homFugacities}(b) 
and~\ref{fig_homFugacities}(d)].
For comparison, solid and dashed lines in Fig.~\ref{fig_homFugacities} show
the fugacities $z_1$ and $z_2$, respectively, for the corresponding
non-interacting systems.
Even for the non-interacting system, the fugacities $z_1$ and $z_2$ are not
symmetric with respect to $P=0$; this is a direct consequence of the
$\kappa^{3/2}$ factor in Eq.~\eqref{eq_homN2}.
For $-1 < P < 1$, the fugacities for the non-interacting systems lie above
the respective fugacities for the unitary systems.
For $\kappa > 1$, this implies that the unitary system has a smaller 
equilibrium polarization $P_{eq}$
than the non-interacting system for the same $T$ and $\kappa$.
Intuitively, this makes sense since we expect that the attractive interactions,
at least for mass ratios not too much larger than $1$, push the system
toward an equal mixture of heavy and light particles.

\begin{figure}
\vspace*{+1.5cm}
\includegraphics[angle=0,width=70mm]{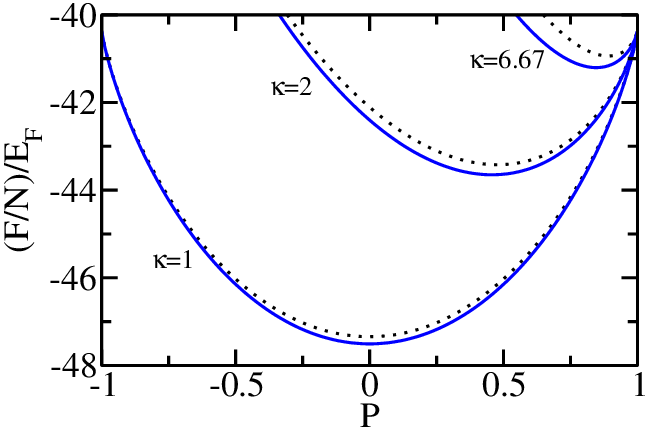}
\vspace*{0.1cm}
\caption{(Color online) Free energy per particle $F/N$ of the homogeneous
system as a function 
of the polarization $P$ for $T/T_F=10$ and mass ratios $\kappa=1,2$, and 6.67.
The curves are grouped in sets of two, each labeled by their corresponding
mass ratio.
The solid lines are calculated using the virial equation of state at unitarity.
For comparison, the dotted lines show $F/N$ for the non-interacting system.
}\label{fig_homF_vs_P}
\end{figure}

Figure~\ref{fig_homF_vs_P}
shows the free energy per particle $F/N$ as a function of the polarization $P$
for $T/T_F=10$ and three different mass ratios $\kappa$.
The solid lines are calculated using the virial equation of state
of the unitary system.
For comparison, dotted lines show $F/N$ for the non-interacting system.
Since the free energy is a concave function with respect to $P$
for fixed $N$ and $T$, the equilibrium polarizations 
are stable minima.

To better understand the interplay of the second and third order virial 
coefficients for the homogeneous system, we calculate 
$z_{eq}$ analogously to Eq.~\eqref{eq_zeq} and find
$P_{eq}$ for the homogeneous system to be
\begin{align}
\label{eq_Peq}
P_{eq} = & \frac{N\lambda_{m_1}^3}{V}\Bigg(-\mbox{Li}_{3/2}(-z_{eq})
\left[1-\frac{1}{\kappa^{3/2}}\right] \nonumber \\
& + [b_{2,1,\rm{hom}}-b_{1,2,\rm{hom}}]z_{eq}^3\Bigg).
\end{align}
Eliminating the polylogarithm function, $P_{eq}$ can be rewritten as
\begin{align}
\label{eq_PeqOpt}
P_{eq} & = \frac{\kappa^{3/2}-1}{\kappa^{3/2}+1} - b_{1,1,\rm{hom}}
\frac{\kappa^{3/2}-1}{\kappa^{3/2}+1} \frac{2V}{N\lambda^3_{m_1}} z_{eq}^2 + \\
& + \Bigg[ b_{1,2,\rm{hom}}^{(0)} \frac{1-2\kappa^{3/2}}{\kappa^{3/2}+1}
+ b_{2,1,\rm{hom}}^{(0)}\frac{2-\kappa^{3/2}}{\kappa^{3/2}+1} \Bigg] 
\frac{2V}{N\lambda^3_{m_1}} z_{eq}^3. \nonumber
\end{align}
As expected, the $z_{eq}^0$ term depends only on the mass ratio $\kappa$
and is greater than zero for $\kappa>1$; in the limit 
$\kappa\rightarrow \infty$, the non-interacting 
system prefers to have only heavy particles, i.e., $P_{eq} \rightarrow 1$.
The $z_{eq}^2$ term of the unitary system, in contrast, 
is negative for all $\kappa>1$. 
This implies that the second order virial coefficient of the unitary system
acts to decrease $P_{eq}$ compared to the $P_{eq}$ of the non-interacting
system.
Interestingly, the $z_{eq}^3$ term of the unitary system is positive for
$\kappa < 5.40$ and negative for $\kappa>5.40$, independent of temperature.
This implies that the third order virial coefficients act to increase
(decrease)
$P_{eq}$ for $1 < \kappa < 5.40$  ($\kappa > 5.40$) compared to the $P_{eq}$
calculated using the virial equation of state up to second order.

\begin{figure}
\vspace*{+1.5cm}
\includegraphics[angle=0,width=70mm]{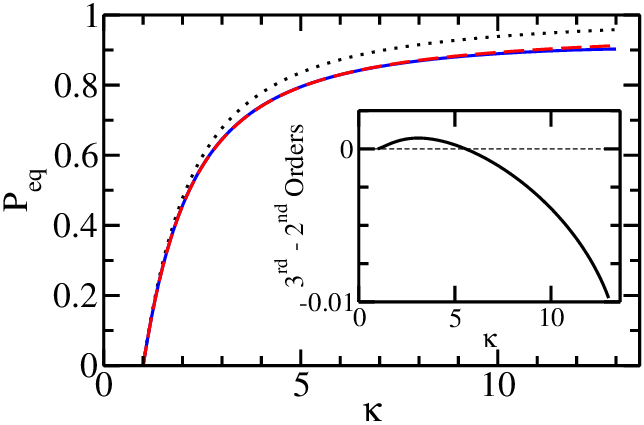}
\vspace*{0.1cm}
\caption{(Color online) Equilibrium polarization $P_{eq}$ of the
homogeneous system at unitarity as a function of
the mass ratio $\kappa$ for $T/T_F=10$.  
The dashed and solid lines show $P_{eq}$ calculated using the virial
equation of state up to second and third order, respectively.
The dashed and solid lines are nearly indistinguishable on the scale shown.
For comparison, the dotted line shows $P_{eq}$ for the
non-interacting system.
The solid line in the inset shows the difference between the $P_{eq}$'s 
calculated using the virial equation of state up to the 
third and second orders (the dashed line shows zero as a reference).
}\label{fig_homFeqA}
\end{figure}

Figure~\ref{fig_homFeqA}
shows the equilibrium polarization $P_{eq}$ as a function of $\kappa$
for $T/T_F=10$.
The dashed and solid lines show $P_{eq}$ calculated using the virial equation
of state of the unitary system up to second and third order, respectively.
The second and third orders are nearly indistinguishable on the scale shown.
For comparison, the dotted line shows $P_{eq}$ for the non-interacting system.
The inset of Fig.~\ref{fig_homFeqA} shows the difference of $P_{eq}$
calculated up to the second order subtracted from $P_{eq}$ calculated up to the
third order. Note that the zero crossing occurs at $\kappa=5.56$, 
and not at $\kappa=5.40$, since the equilibrium fugacity determined from the
virial equation of state up to third order is slightly smaller than that
determined from the virial equation of state up to second order.

\begin{figure}
\vspace*{+1.5cm}
\includegraphics[angle=0,width=70mm]{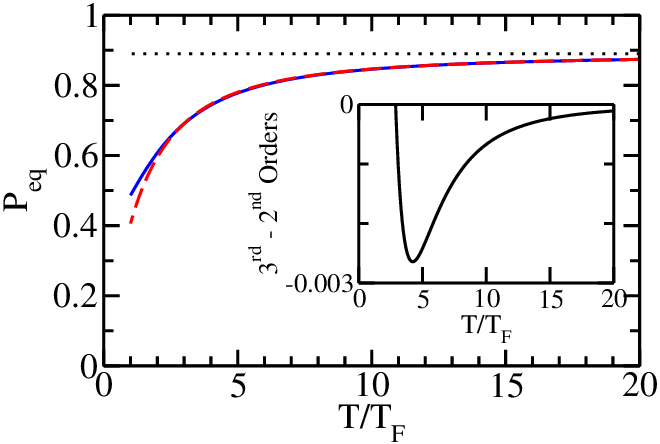}
\vspace*{0.1cm}
\caption{(Color online) Equilibrium polarization $P_{eq}$ of the
homogeneous system at unitarity as a function of
the temperature $T$ for $\kappa=6.67$.  
The dashed and solid lines show $P_{eq}$ calculated using the virial
equation of state up to second and third order, respectively.
The dashed and solid lines are nearly indistinguishable on the scale shown.
For comparison, the dotted line shows $P_{eq}$ for the
non-interacting system.
The solid line in the inset shows the difference between the $P_{eq}$'s 
calculated using the virial equation of state up to the 
third and second orders.
}\label{fig_homFeqB}
\end{figure}

Figure~\ref{fig_homFeqB}
shows the equilibrium polarization 
$P_{eq}$ as a function of $T$ for $\kappa=6.67$. The dashed and solid lines
show $P_{eq}$ at unitarity calculated using the virial equation of state
up to second and third order, respectively.
The second and third orders are nearly indistinguishable on the scale shown.
For comparison, the dotted line shows $P_{eq}$ for the
non-interacting system;
it is independent of $T$ and equals $0.89$.
The inset of Fig.~\ref{fig_homFeqB} shows the difference of $P_{eq}$
at unitarity
calculated up to the second order subtracted from $P_{eq}$ 
at unitarity calculated up to the third order.
Since the mass ratio used is larger than $5.40$, the third order correction 
acts to further decrease $P_{eq}$ for $T/T_F > 5$.
Note that the virial equation of state is likely unreliable 
for $T/T_F \lesssim 6.67$.
As the temperature decreases, interactions become more important
and the deviations between $P_{eq}$ for the unitary and the non-interacting
system increase.

\section{Conclusions}
\label{sec_conclusion}
This paper considers the high temperature virial equation of state of
unequal-mass two-component Fermi gases at unitarity.
For the trapped system with equal frequencies and $\kappa>1$, we found that
the equilibrium polarization $P_{eq}$ at unitarity is relatively small for
the temperatures considered.
The small deviations from $P_{eq}=0$ are introduced by the third order virial
coefficients.
For the homogeneous system, in contrast, we found comparitively large $P_{eq}$
for $\kappa>1$.
The value of $P_{eq}$ is determined by an intricate interplay between the 
second and third order virial coefficients.
For the cases investigated, $P_{eq}$ increases with decreasing temperature for
the trapped system while $P_{eq}$ decreases with temperature for the
homogeneous system.
The fourth-order virial coefficients $b_{3,1}$ and $b_{2,2}$ for unequal-mass
systems are presently unknown, preventing us to estimate the importance
of higher-order terms in the virial equation of state.

At the temperatures considered, the curvature of the free energy
is positive for all polarizations, thus 
implying the absence of any first-order phase transition. 
If the unequal-mass system is prepared with a polarization $P$
different from $P_{eq}$, then the free energy lies above the value
it would have for $P_{eq}$.  This suggests that the system
prepared with $P \ne P_{eq}$ may phase separate when the temperature
falls below a certain critical value.

In future work, it will be interesting to investigate the high temperature
thermodynamics of unequal mass systems when the light and heavy species are
confined by harmonic potentials with different frequencies via the local
density approximation.
Moreover, the treatment of cylindrically symmetric traps will be relevant for
ongoing experiments.
Other future extensions of our work include the treatment of unequal mass
systems away from unitarity and at lower temperatures.

\section{Acknowledgements}
Support by the NSF through grant PHY-0855332
is gratefully acknowledged.
We also acknowledge discussions with Stefano Giorgini on the use of the local
density approximation.

\appendix
\section{Thermodynamics of the trapped single-component fermi gas}
\label{sec_thermo1}

This appendix studies the virial expansion
of the single-component Fermi gas in a harmonic trap. 
The results for the single-component Fermi gas serve as a reference for the
trapped $s$-wave interacting two-component Fermi gas.
We first derive compact expressions for the virial coefficients, and
then discuss convergence properties of the virial expansion.

Equation \eqref{eq_singCompFGa} of the main text
defines the expansion coefficients $b_n$;
throughout this appendix, we replace $n_1$ by $n$ and use a single subscript 
$n$ to denote the virial coefficients $b_n$ and the canonical
partition functions $Q_n$.
Partial derivatives of the grand canonical potential $\Omega^{(1)}$ determine
the various thermodynamic quantities.
The number of particles $N$ of the single-component Fermi gas, e.g., can be
expressed as a derivative with respect to the chemical potential $\mu$,
\begin{align}
\label{eq_singCompN}
N=-\frac{\partial \Omega^{(1)}}{\partial \mu} = 
Q_1 \sum_{n=1}^{\infty} n b_n z^n.
\end{align}
To determine the coefficients $b_n$, we use that 
the population $N(E_j)$ of a state with a given single-particle energy 
$E_j$ is given by the Fermi-Dirac distribution,
\begin{align}
\label{eq_FermiDirac}
N(E_j) = \frac{1}{e^{(E_j - \mu)/(k_B T)} + 1},
\end{align}
with the constraint that the total number of particles $N$ is given by
\begin{align}
\label{eq_Ntotal}
N=\sum_{j=1}^{\infty} N(E_j).
\end{align}
The single-particle energies of the three-dimensional 
harmonic oscillator are given by 
$E_j = (j+1/2)\hbar\omega$, $j=1,2,3\ldots$, and have a degeneracy of 
$j (j+1)/2$. 
Using the single-particle 
energies and their degeneracies in Eqs.~
(\ref{eq_FermiDirac}) and (\ref{eq_Ntotal}), we find a relationship between
$N$, $\mu$, and $T$,
\begin{align}
\label{eq_NmuT}
N = \sum_{j=1}^{\infty} \frac{j (j+1)}{2} \left[e^{\left(j+\tfrac{1}{2}-\tfrac{\mu}{\hbar\omega}\right) \tilde{\omega}} + 1\right]^{-1}.
\end{align}

To compare Eqs.~(\ref{eq_NmuT}) and (\ref{eq_singCompN}), we Taylor expand 
Eq. (\ref{eq_NmuT}) for small $z$, $z=\exp[\mu / (k_B T)]$,
\begin{align}
\label{eq_NmuTsmallZ}
N = \sum_{j=1}^{\infty} \frac{j (j+1)}{2} \sum_{n=1}^{\infty} (-1)^{n+1} e^{-\left(j+\tfrac{1}{2}\right) \tilde{\omega} n} z^n.
\end{align}
The sum over $j$ can be done analytically.
Pulling out a factor of $Q_1$, we obtain
\begin{align}
\label{eq_NmuTend}
N = Q_1 \sum_{n=1}^{\infty} (-1)^{n+1} \; 
e^{\tfrac{3}{2}(n-1)\tilde{\omega}}
\left(\frac{e^{\tilde{\omega}}-1}{e^{n\tilde{\omega}}-1}\right)^3 z^n.
\end{align}
Comparing Eqs.~(\ref{eq_NmuTend}) and (\ref{eq_singCompN}), the expansion
coefficients $b_n$ can be read off,
\begin{align}
\label{eq_singCompVirial}
b_n = \frac{(-1)^{n+1}}{n} \; 
e^{\tfrac{3}{2}(n-1)\tilde{\omega}}
\left(\frac{e^{\tilde{\omega}}-1}{e^{n\tilde{\omega}}-1}\right)^3.
\end{align}

\begin{figure}
\vspace*{+1.5cm}
\includegraphics[angle=0,width=70mm]{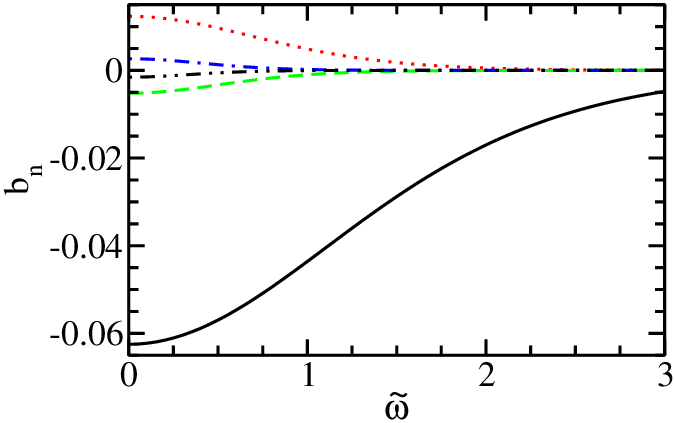}
\vspace*{0.1cm}
\caption{(Color online) Virial coefficients $b_n$ of 
the single-component harmonically trapped 
Fermi gas as a function of $\tilde{\omega}$. 
Solid, dotted, dashed, dash-dotted and
dash-dot-dotted lines show $b_2,b_3,b_4,b_5$ and $b_6$, respectively.
}\label{fig_bn}
\end{figure}

Figure \ref{fig_bn}
shows the first five expansion coefficients as a function of $\tilde{\omega}$.
In the low temperature limit, the $b_n$ decay exponentially as 
$\exp[-3(n-1)\tilde{\omega}/2]$.
The sign of $b_n$ is given by $(-1)^{n+1}$ for all $\tilde{\omega}$ and
the $b_n$ monotonically decrease in magnitude as $\tilde{\omega}$ increases.
The coefficients are even functions in $\tilde{\omega}$, i.e.,
$b_n$ remains unchanged as $\tilde{\omega}$ is replaced by $-\tilde{\omega}$.
Expanding Eq.~(\ref{eq_singCompVirial}) for large $T$ (small $\tilde{\omega}$),
we find~\cite{notation4}
\begin{align}
\label{eq_singCompVirialExpansion}
b_n \approx & \frac{ (-1)^{n+1}}{n^4}\biggr[ 1 
- \frac{1}{8}(n^2-1)\tilde{\omega}^2 \nonumber \\
& + \frac{1}{1920}(17n^4-30n^2+13)\tilde{\omega}^4 
+ {\cal{O}}(\tilde{\omega}^6) \biggr].
\end{align}
In the limit $\tilde{\omega}\rightarrow0$ we have ``universal'' 
physics in the sense that the virial coefficients $b_n$ approach constants,
$b_n \rightarrow b_n^{(0)} $ as $\tilde{\omega}\rightarrow 0$, 
where $b_n^{(0)} =(-1)^{n+1}/n^4$~\cite{LiuPRL2009}.

\begin{figure}
\vspace*{+1.5cm}
\includegraphics[angle=0,width=70mm]{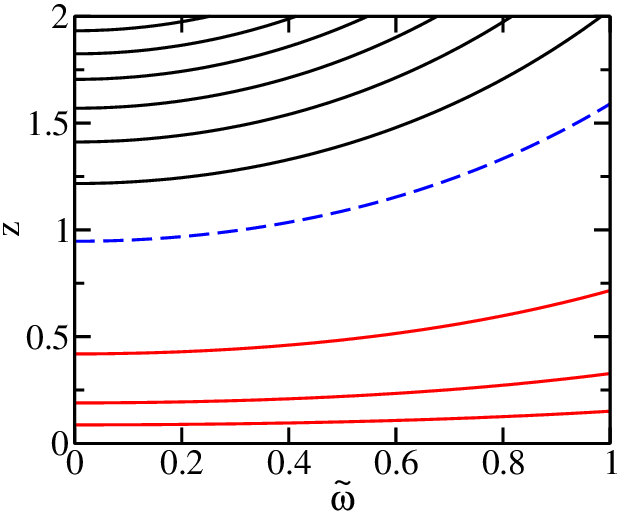}
\vspace*{0.1cm}
\caption{(Color online) Contour plot of the percent difference 
$100(z + 2 b_2 z^2 + 3 b_3 z^3 - N/Q_1)/(N/Q_1)$.
The dashed line shows the percent difference of 1.
The solid lines below the dashed line show the percent difference in
steps of powers of 10 down to a value of 0.001, 
while solid lines above the dashed line show the percent difference in 
steps of 1 up to a value of 7.
We consider the region where the percent difference is less than 1
to be well converged.
}\label{fig_convergence}
\end{figure}

Figure~\ref{fig_convergence}
shows a contour plot of
the percent difference between $z + 2 b_2 z^2 + 3 b_3 z^3$ and $N/Q_1$
[see Eq.~\eqref{eq_singCompN}].
The deviation between the exact expressions and the virial expansion
is smaller than about 1\% for $z \le 1$ and is below 5\% for $z \approx 1.5$
for the range of temperatures shown.
Thus, the virial expansion up to the third order provides a qualitatively 
correct description of the high-temperature physics for $z \alt 1.5$.
Even though we expanded around small $z$,
Eq.~(\ref{eq_singCompN}) is an analytic function of $z$ 
that can be continued along the positive real axis for all $z \ge 1$~\cite{KahnUhlenbeckPhysica1938}.
For the two-component Fermi gas with pairwise interactions, in contrast,
the convergence radius of the virial equation of state,
to the best of our knowledge, is not known.

Lastly, one can use Eq.~(\ref{eq_singCompVirial})
to calculate the $Q_n$ for $n>1$.
A compact expression for the $Q_n$, $n>1$, can be found in the literature~\cite{KahnUhlenbeckPhysica1938,HuangStatMech,MayerMayerStatMech},
\begin{align}
\label{eq_QnAll}
Q_n = \sum_{\{ \vec{m} \} } \prod_{i=1}^n \frac{1}{m_i!} (Q_1 b_i)^{m_i},
\end{align}
where $\{ \vec{m} \}$ 
represents all possible sets of $n$ non-negative integers
$\{ m_1,\ldots,m_n \}$ that fullfill the constraint
\begin{align}
\label{eq_QnAllb}
\sum_{i=1}^n i m_i = n.
\end{align}
Each set describes a unique way to divide a system of $n$ particles
into smaller clusters of $m_i$ particles.
For $n=3$, e.g., we have the sets $\{ 3,0,0 \}$, $\{ 1,1,0 \}$, 
and $\{ 0,0,1\}$, i.e., the three-body system can be thought of as consisting
of three monomers, of one monomer and one dimer, or of one trimer.
For completeness, we report the expressions for the first few $Q_n$,
\begin{align}
\label{eq_q1}
Q_1 = & \sum_{n=0}^{\infty} \sum_{l=0}^{\infty} (2l+1)e^{-(2n+l+3/2)\tilde{\omega}}
= \frac{e^{3 \tilde{\omega}/2}}{(e^{\tilde{\omega}}-1)^3}, \\
\label{eq_Qn2}
Q_2 = & \; b_2 Q_1 + \tfrac{1}{2}Q_1^2, \\
\label{eq_Qn3}
Q_3 = & \; b_3 Q_1 + b_2 Q_1^2 + \tfrac{1}{6} Q_1^3,
\end{align}
and
\begin{align}
\label{eq_Qn4}
Q_4 = & \; b_4 Q_1 + b_3 Q_1^2 + \tfrac{1}{2} b_2^2 Q_1^2 
+ \tfrac{1}{2} b_2 Q_1^3 + \tfrac{1}{24} Q_1^4,
\end{align}
where we have used that $b_1=1$ [see Eq.~\eqref{eq_singCompVirialExpansion}].


\begin{thebibliography}{10}
\bibitem{notation2}
For the harmonically trapped gas, where atoms of species 1 and 2 feel the
same trapping frequency, one finds $Q_{1,0}=Q_{0,1}$ (and more generally,
$Q_{n_1,0}=Q_{0,n_2}$ for $n_1=n_2$) and Eqs.~\eqref{eq_v11}-\eqref{eq_v21} 
can be simplified.
For the homogeneous system, in contrast, $Q_{1,0}$ and $Q_{0,1}$ differ if
$m_1 \ne m_2$.  Our definition of the virial coefficients is chosen such
that it applies to the trapped and homogeneous systems.
\bibitem{notation3}
As can be seen from the expression for $T_F$, the Fermi temperature of the 
trapped
non-interacting system is independent of the mass of the constituents and our 
reference Fermi temperature can alternatively be interpreted as the Fermi
temperature of the single-component Fermi gas with $N/2$ particles of mass 
$m_2$.
For the homogeneous system (see Sec.~\ref{sec_thermo3}), in contrast,
the Fermi temperature depends on the mass of the constituents and we define
$T_F$ in terms of $m_1$ in that case.
\end{thebibliography}
\end{document}


\title{Auxiliary material for ``Thermodynamics of the trapped
two-component Fermi gas with unequal masses at unitarity''}

\author{K.~M. Daily}
\affiliation{Department of Physics and Astronomy,
Washington State University,
  Pullman, Washington 99164-2814, USA}
\author{D. Blume}
\affiliation{Department of Physics and Astronomy,
Washington State University,
  Pullman, Washington 99164-2814, USA}

\date{\today}
\maketitle

This material tabulates the expansion coefficients $b_{1,2}^{(n)}$ 
and $b_{2,1}^{(n)}$ ($n=0,2,4,$ and 6) of
the two-component unequal-mass Fermi gas 
in a harmonic trapping potential at unitarity.
The calculation of the virial coefficients $b_{1,2}$ and $b_{2,1}$ 
requires the eigenvalues $s_{l,\nu}$ [see Eqs.~(21)-(24) of the main text].
We calculate the first 10,000 $s_{l,\nu}$ 
for each $l$, $l=0-50$, and each mass ratio considered.
To obtain the accuracy reported in Tables~\ref{table_highTa} and~\ref{table_highTb},
we must maintain a higher working precision than machine precision.
We use Mathematica to determine the $s_{l,\nu}$ with an accuracy of about
one part in $10^{80}$ (most likely, a notably reduced precision would work 
just as well).
To obtain the $b_{1,2}^{(n)}$ and $b_{2,1}^{(n)}$, we Taylor expand Eqs.~(23) 
and~(24) of the main text to high order at a finite but small $\tilde{\omega}$.
The $\tilde{\omega}$ is chosen such that $b_{1,2}$ and $b_{2,1}$ are converged
at that value; note, since we use a cutoff in $\nu$ and $l$, we cannot Taylor
expand around $\tilde{\omega}=0$. To determine the expansion coefficients for
$\tilde{\omega}=0$, we rewrite the Taylor expanded series in the form of
Eq.~(25) of the main text.
For all values listed in Tables~\ref{table_highTa} and~\ref{table_highTb}, 
the error is indicated by the value in parentheses.

\begin{table}
\caption{High temperature expansion coefficients 
$b_{1,2}^{(0)}$, $b_{1,2}^{(2)}$, $b_{1,2}^{(4)}$, and $b_{1,2}^{(6)}$
for one heavy fermion and two light fermions for various mass ratios $\kappa$.
The odd order coefficients are zero to the calculated accuracy.
For the mass ratio of 13.607, we use 67907169/49906069.
}
\begin{ruledtabular}
\begin{tabular}{c|r@{.\hspace{-6mm}}lr@{.\hspace{-6mm}}lr@{.\hspace{-6mm}}lr@{.\hspace{-6mm}}l}
$\kappa$ & \multicolumn{2}{c}{$b_{1,2}^{(0)}$} & \multicolumn{2}{c}{$b_{1,2}^{(2)}$} & \multicolumn{2}{c}{$b_{1,2}^{(4)}$} & \multicolumn{2}{c}{$b_{1,2}^{(6)}$} \\  \hline
$\infty$ & \multicolumn{2}{c}{-1/8} & \multicolumn{2}{c}{3/32} & \multicolumn{2}{c}{-3/64} & \multicolumn{2}{c}{157/7680} \\ 
$\approx13.607$ & -0&1148983967419143(1) & 0&0822846200355016(1) & -0&0387505068487700(1) & 0&0157412555073196(1) \\  
12 & -0&1137494461215969(1) & 0&0810220572507766(1) & -0&0378883247633289(1) & 0&0152623106544741(1) \\  
10 & -0&1118930949742869(1) & 0&0790006163965818(1) & -0&0365219657732420(1) & 0&0145117031903192(1) \\  
8 & -0&1092998807903862(1) & 0&0762155583916566(1) & -0&0346683837395572(1) & 0&0135104326391047(1) \\  
6.67 & -0&1069167194627094(1) & 0&0736967168785517(1) & -0&0330216423362948(1) & 0&0126379464141623(1) \\  
6 & -0&1054166227589069(1) & 0&0721315542690726(1) & -0&0320129640005759(1) & 0&0121117516797285(1) \\  
4 & -0&0989274431349124(1) & 0&0655479723141608(1) & -0&0278982228513893(1) & 0&0100342277821471(1) \\  
2 & -0&0854120042021242(1) & 0&0528794100025025(1) & -0&0206256339544718(1) & 0&0066746031546180(1) \\  
1 & -0&0683396093112849(1) & 0&0388697595336693(1) & -0&0136736950010533(1) & 0&0039221710096865(1)
\end{tabular}
\end{ruledtabular}
\label{table_highTa}
\end{table}

\begin{table}
\caption{
High temperature expansion coefficients 
$b_{2,1}^{(0)}$, $b_{2,1}^{(2)}$, $b_{2,1}^{(4)}$, and $b_{2,1}^{(6)}$
for two heavy fermions and one light fermion for various mass ratios $\kappa$.
The odd order coefficients are zero to the calculated accuracy.
For the mass ratio of 13.607, we use 67907169/49906069.
}
\begin{ruledtabular}
\begin{tabular}{c|r@{.\hspace{-6mm}}lr@{.\hspace{-6mm}}lr@{.\hspace{-6mm}}lr@{.\hspace{-6mm}}l}
$\kappa$ & \multicolumn{2}{c}{$b_{2,1}^{(0)}$} & \multicolumn{2}{c}{$b_{2,1}^{(2)}$} & \multicolumn{2}{c}{$b_{2,1}^{(4)}$} & \multicolumn{2}{c}{$b_{2,1}^{(6)}$} \\  \hline
1      & -0&0683396093112849(1) & 0&0388697595336693(1) & -0&0136736950010533(1) & 0&0039221710096865(1) \\  
2      & -0&0392970167749193(1) & 0&0168593086798205(1) & -0&0035955339421578(1) & 0&0002304139076276(1) \\  
3      & -0&0043665570747450(1) & -0&0113457745372200(1) & 0&010983366385180(9) & -0&0062006157599(2) \\  
4      & 0&039557237585138(3) & -0&0500118892685(1) & 0&034191104164(3) & -0&01875475605(4) \\  
5      & 0&09356648286595(9) & -0&101149373357(4) & 0&0692162284(1) & -0&041313819(2) \\  
6      & 0&158595249651(2) & -0&16632882903(7) & 0&119273264(2) & -0&07878135(4) \\  
6.67   & 0&208903183724(6) & -0&2185938559(4) & 0&162900512(9) & -0&1152027(2) \\  
7      & 0&23583203496(2) & -0&2470369251(7) & 0&18775226(2) & -0&1372414(4) \\  
8      & 0&32697976919(7) & -0&344820636(4) & 0&2782550(2) & -0&224063(3) \\  
9      & 0&4346241735(3) & -0&46141279(2) & 0&3946117(6) & -0&34799(2) \\  
10     & 0&562958249(2) & -0&59892409(7) & 0&540902(3) & -0&51922(5) \\  
11     & 0&719509561(4) & -0&7602484(2) & 0&721507(7) & -0&7495(2) \\  
11.5   & 0&812728781(5) & -0&8511988(4) & 0&82616(2) & -0&8909(3) \\  
12     & 0&920430123(8) & -0&9502037(6) & 0&94125(2) & -1&0522(5) \\  
12.25  & 0&98168666(1) & -1&0031950(7) & 1&00295(3) & -1&1410(6) \\  
12.5   & 1&04968722(2) & -1&0589100(9) & 1&06756(4) & -1&2350(8) \\  
12.75  & 1&12681250(2) & -1&117846(2) & 1&13521(4) & -1&336(1) \\  
13     & 1&21738420(2) & -1&180858(2) & 1&20609(5) & -1&443(2) \\  
13.1   & 1&25917000(2) & -1&207546(2) & 1&23540(6) & -1&487(2) \\  
13.2   & 1&30557689(3) & -1&235350(2) & 1&26530(6) & -1&533(2) \\  
13.3   & 1&35844252(3) & -1&264602(2) & 1&29590(7) & -1&579(2) \\  
13.4   & 1&42130482(3) & -1&295930(2) & 1&32710(7) & -1&627(2) \\  
13.5   & 1&50304784(3) & -1&330880(2) & 1&35930(8) & -1&676(2) \\  
13.51  & 1&51310826(3) & -1&334720(2) & 1&36250(8) & -1&681(2) \\  
13.52  & 1&52371358(3) & -1&338657(2) & 1&36580(8) & -1&686(2) \\  
13.53  & 1&53496097(3) & -1&342707(2) & 1&36920(8) & -1&691(2) \\  
13.54  & 1&54698040(3) & -1&346893(2) & 1&37250(8) & -1&696(2) \\  
13.55  & 1&55995276(3) & -1&351246(2) & 1&37590(8) & -1&701(2) \\  
13.56  & 1&57414297(3) & -1&355809(3) & 1&37930(8) & -1&706(2) \\  
13.57  & 1&58996766(3) & -1&360653(3) & 1&38270(8) & -1&711(3) \\  
13.58  & 1&60815453(3) & -1&365900(3) & 1&38620(8) & -1&716(3) \\  
13.59  & 1&63021654(3) & -1&371803(3) & 1&38970(8) & -1&721(3) \\  
13.595 & 1&64379580(3) & -1&375185(3) & 1&39160(8) & -1&724(3) \\  
13.596 & 1&64683813(3) & -1&375916(3) & 1&39190(8) & -1&724(3) \\  
13.597 & 1&65002366(3) & -1&376671(3) & 1&39230(8) & -1&725(3) \\  
13.598 & 1&65337456(3) & -1&377454(3) & 1&39270(8) & -1&725(3) \\  
13.599 & 1&65691940(3) & -1&378269(3) & 1&39310(8) & -1&726(3) \\  
13.6   & 1&66069607(3) & -1&379124(3) & 1&39350(8) & -1&726(3) \\  
13.601 & 1&66475668(3) & -1&380025(3) & 1&39390(8) & -1&727(3) \\  
13.602 & 1&66917640(3) & -1&380987(3) & 1&39430(8) & -1&727(3) \\  
13.603 & 1&67407090(3) & -1&382028(3) & 1&39470(8) & -1&728(3) \\  
13.604 & 1&67963563(3) & -1&383182(3) & 1&39510(8) & -1&728(3) \\  
13.605 & 1&68625428(3) & -1&384511(3) & 1&39550(8) & -1&729(3) \\  
13.606 & 1&69493193(3) & -1&386183(3) & 1&39600(8) & -1&729(3) \\  
$\approx13.607$ & 1&71529815(3) & -1&389797(3) & 1&39670(8) & -1&730(3) \\  
\end{tabular}
\end{ruledtabular}
\label{table_highTb}
\end{table}